\documentclass{elsarticle}

\usepackage{booktabs}

\newcommand{\td}{\mathrm{d}}
\newcommand{\te}{\mathrm{e}}
\newcommand{\ti}{\mathrm{i}}

\newcommand{\ua}{\uparrow}
\newcommand{\da}{\downarrow}

\usepackage{graphicx}
\graphicspath{{./}{./img/}{../img/}}

\usepackage[bookmarks]{hyperref}
\usepackage{enumerate}

\usepackage{amsmath}
\usepackage{amssymb}

\renewcommand{\today}{July 6, 2015}
\journal{Annals of Physics}

\begin{document}

 \begin{frontmatter}
 \title{Criteria for the absence of quantum fluctuations after spontaneous symmetry breaking}
 
\author[cems]{Aron~J.~Beekman}
\ead{beekman.aronjonathan@nims.go.jp}

\address[cems]{RIKEN Center for Emergent Matter Science (CEMS), Wako 351-0198, Japan\\
present address: NIMS Advanced Key Technologies Division, Tsukuba, Japan}

 \begin{abstract}
The lowest-energy state of a macroscopic system in which symmetry is spontaneously broken, is a very stable wavepacket centered around a spontaneously chosen, classical direction in symmetry space. However, for a Heisenberg ferromagnet the quantum groundstate is exactly the classical groundstate, there are no quantum fluctuations. This coincides with seven exceptional properties of the ferromagnet, including spontaneous time-reversal symmetry breaking, a reduced number of Nambu--Goldstone modes and the absence of a thin spectrum (Anderson tower of states). Recent discoveries of other non-relativistic systems with fewer Nambu--Goldstone modes suggest these specialties apply there as well. I establish precise criteria for the absence of quantum fluctuations and all the other features. In particular, it is not sufficient that the order parameter operator commute with the Hamiltonian. It leads to a measurably larger coherence time of superpositions in small but macroscopic systems.
 \end{abstract}
 

\end{frontmatter}

\section{Introduction}\label{sec:Introduction}
The Heisenberg ferromagnet has always been an eccentric duckling in the flock of spontaneous symmetry breaking (SSB) states consisting of antiferromagnets, crystals, superfluids, chiral SSB, the Standard Model and many others. This is only exacerbated by being one of the earliest and simplest models demonstrating SSB, used as the archetype in a large portion of the literature. Perhaps because much of its physics can be understood by undergraduate level calculations, have its peculiarities never been put in a larger perspective. Still the subtleties are intricate enough to have sparked debates between the greatest of minds in the past century~\cite{AndersonLangackerMann90,PeierlsKaplanAnderson91}.

Why is this state different from all other states? We talk about the following observations, clarified below: 

\begin{enumerate}[(i)]
 \item the order parameter operator commutes with the Hamiltonian, is therefore a symmetry generator and is conserved in time; 
 \item two broken symmetry generators correspond to a single, quadratically dispersing Nambu--Goldstone (NG) mode;
 \item the classical groundstate is an exact eigenstate of the Hamiltonian, there are no quantum fluctuations;
 \item the raising operator, a root generator of the symmetry algebra, annihilates the groundstate, even locally (the spin of the maximally polarized state cannot be increased);
 \item there is no {\em thin spectrum} or {\em tower of states} of nearly vanishing energy just above the groundstate;
 \item the groundstate is an eigenstate of the unbroken symmetry generator with non-zero eigenvalue;
 \item time-reversal symmetry is spontaneously broken.
\end{enumerate}

Arguably the most important of these features are (i) and (ii): the low-energy spectrum of NG modes is different from what one would expect based on the relativistic Goldstone theorem. This issue had been recognized early on, \cite{Lange65,Lange66,Wagner66}, and later generalized \cite{NielsenChadha76} to systems other than the ferromagnet, but has basically been solved only in the last ten years or so~\cite{SchaferEtAl01,Nambu04,Brauner10,WatanabeBrauner11,WatanabeMurayama12,Hidaka13,WatanabeMurayama14r}: whenever the order parameter operator $Q^k$, that obtains a non-zero expectation value $\langle Q^k \rangle$ in the symmetry-broken state, is one of the symmetry generators itself---called a {\em finite Noether charge density}---then any two spontaneously broken generators $Q^i,Q^j$ that contain this operator in their commutation relation $[Q^i,Q^j] = \sum_k f_{ijk} Q^k$ will in fact excite {\em the same} NG mode. That mode will have a different, in general quadratic, dispersion relation (a more precise statement will be given below). Therefore (i) implies (ii). For the ferromagnet with magnetization along the $z$-axis, the spin rotation operator $S^z$ obtains a finite Noether charge density, while $S^x$ and $S^y$ are spontaneously broken and excite the same single spin wave (magnon) with quadratic dispersion. Such modes have been called type-B NG modes, as opposed to the `regular', linearly dispersing, type-A NG modes. 

In a parallel development, several states of matter with broken charge densities and/or quadratically dispersing NG modes other than the ferromagnet have been identified. For instance in spinor Bose--Einstein condensates (BEC)~\cite{KawaguchiUeda12}, kaon condensates in quantum chromodynamics~\cite{SchaferEtAl01} and Tkachenko modes in superfluid vortices~\cite{WatanabeMurayama13}. The question of whether the other special properties of the ferromagnet (iii)--(vii) generalize to such systems arises naturally. 

Here I will establish precise criteria for the relations between each of the properties (i)--(vii). I will focus in particular on quantum fluctuations, properties (iii)--(v). When a continuous symmetry is spontaneously broken, there is a continuous manifold of degenerate classical groundstates. In the quantum case, any superposition of these states will be a valid groundstate as well, but tiny external perturbations will favor one particular classical state over all the others. At this point that may seem obvious, but these classical groundstates are almost never eigenstates of the quantum Hamiltonian, which implies unitary time evolution would bring one to a state different from the classical state. This deviation from the classical groundstate is known as {\em quantum fluctuations} although there is actually no time-dependent behavior, in the same sense as there are no particles and anti-particles ``popping into and out of existence'' in the QED vacuum. It is perhaps the most striking feature of spontaneous symmetry breaking that in almost all cases the actual quantum groundstate is very close to a classical groundstate \cite{Anderson84,Weinberg96b,WezelBrink07}; for instance the reader's chair is (for all practical purposes) in a position eigenstate even though its Hamiltonian has translation invariance and therefore its spectrum consists of momentum eigenstates. Assuredly the chair's position does not fluctuate in time.

In many texts the fact that classical states are dressed with quantum fluctuations is glossed over or ignored. Several others, most explicitly by Anderson~\cite{Anderson84}, claim that all of the peculiar properties and in particular the absence of quantum fluctuations of the ferromagnet follow from the fact that its order parameter operator, $S^z$, commutes with the Hamiltonian $H$. The reason cited is that in that case, $S^z$ and $H$ can be simultaneously diagonalized and they share a basis of eigenstates. However, this is not sufficient. Namely, it is only the {\em total} magnetization $S^z = \sum_j S^z_j$ where $j$ runs over lattice sites, that commutes with $H$, while the local magnetization density $S^z_j$ does not. Therefore, even if the total magnetization is conserved, local fluctuations of magnetization that leave the total magnetization constant, could in principle be allowed. I will present below some counterexamples of states with conserved order parameters but nevertheless showing quantum fluctuations. As a mnemonic, the reader can picture a ferrimagnet (see Sec.~\ref{subsec:example Ferrimagnet}) with a tiny imbalance between the magnitude of the spins on the A- and B-sublattice, that thus develops a small magnetization and a ferromagnet-like NG mode spectrum, but appears nevertheless to resemble more closely an antiferromagnet. 

In this light, the absence of quantum fluctuations in the Heisenberg ferromagnet (iii) is not due simply to a conserved order parameter, but rather is a consequence of the maximum polarization of the groundstate (iv). In words, fluctuations that leave the total magnetization invariant necessarily involve raising a spin at some site and lowering a spin at some other site (or rather a wavelike superposition of such processes). But the ferromagnet has maximum polarization at each site: the magnetic quantum number $m$ is equal to the spin $s$ at all sites. Therefore such processes would take one outside of the Hilbert space and are forbidden.   

From a symmetry-group-theoretic point of view, it turns out that identifying a single order parameter operator is not enough to determine whether quantum fluctuations are present or absent. Instead, one should consider all linearly independent operators that could function as order parameter, i.e. that obtain an expectation value in the ordered (symmetry-broken) state. Only if all of them commute with the Hamiltonian, will we be in a maximally polarized state and will it feature an absence of quantum fluctuations. If there is at least one linearly independent, non-commuting order parameter operator, then the state will not be maximally polarized, the massless NG mode will appear together with a massive `partner' mode, and quantum fluctuations away from the classical groundstate are present. This is consistent with the recent classification scheme by Hayata and Hidaka~\cite{HayataHidaka15} (see also~\cite{GongyoKarasawa14}), which categorizes type-B massless NG modes with or without massive (gapped) partner modes, next to type-A massless NG modes.

Property (vi), the fact that an unbroken symmetry generator does not annihilate the broken groundstate, as for instance spin-rotations around the $z$-axis in a ferromagnet with magnetization along the $z$-axis, is automatically satisfied by any finite Noether charge density (any symmetry generator that is also an order parameter operator). In fact, this can be `mended' through redefining the symmetry operator by subtracting the order parameter expectation value. However, this will modify the symmetry Lie algebra relations, and later we will find it useful to retain this expectation value in relation to the highest-weight states (maximally polarized states).

Property (vii), the spontaneous breaking of time-reversal invariance, will be shown to be a completely separate issue: some finite Noether charge densities correspond to time-reversal invariant states, while others will break that symmetry. 

\begin{figure}[t]
\begin{center}
\def\svgwidth{7cm}
\begingroup%
  \makeatletter%
  \setlength{\unitlength}{\svgwidth}%
  \global\let\svgwidth\undefined%
  \global\let\svgscale\undefined%
  \makeatother%
  \begin{picture}(1,0.36869722)%
    \put(0,0){\includegraphics[width=\unitlength]{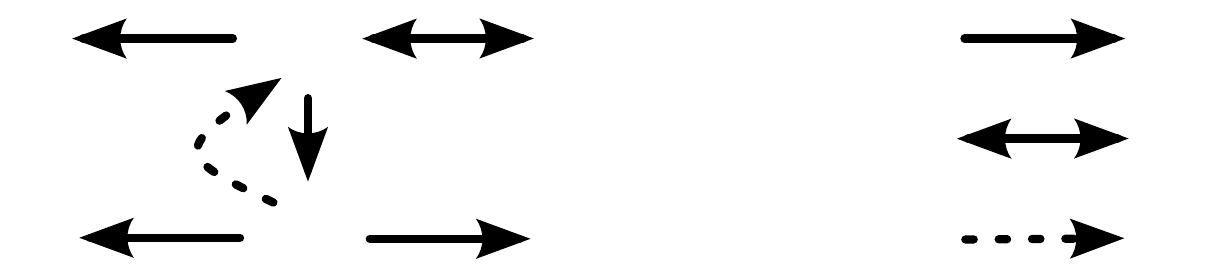}}%
     \put(0.21,0.18){(iii)}%
    \put(0.44,0.18){(iv)}%
    \put(-0.00296365,0.18){(v)}%
    \put(0.57,0.0901246){\fbox{(vii)}}%
    \put(0.23,0.015){(i)}%
    \put(0.44,0.015){(ii)}%
    \put(-.02,0.015){(vi)}%
    \put(0.94,0.18){implies}%
    \put(0.94,0.10){equivalent}%
    \put(0.94,0.015){see text}%
  \end{picture}%
\endgroup%
 \caption{Schematic of implications of properties (i)--(vii). The classical groundstate being an exact eigenstate of the Hamiltonian (iii) is the strongest condition. Only when the total order parameter following from  Eq. 
\eqref{eq:total interpolating field} commutes with the Hamiltonian, (i) implies an exact eigenstate (iii). Time-reversal breaking (vii) is an independent phenomenon.}
\label{fig:FM equivalencies}
\end{center}
\end{figure}
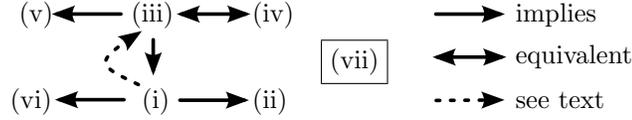

Summarizing, the classical groundstate being an exact eigenstate of the Hamiltonian (iii), or equivalently the groundstate being the maximally polarized state (iv), is the strongest condition, and it implies the other properties in (i)--(vi), see Fig. \ref{fig:FM equivalencies}. Quantum fluctuations are only absent for states which are completely polarized with respect to all broken symmetry generators (the highest-weight state for all broken root generators of the Lie algebra). Otherwise, for systems that have finite Noether charge densities, there will generally be  additional order parameter operators that characterize the symmetry breaking. In that case the type-B NG mode is accompanied by a gapped mode, together excited by the pair of Lie algebra roots.

In this work I will present a reasonably self-contained exposition of known results and a derivation of the claims made in this introduction. To illustrate the matters at hand, in Section \ref{sec:Heisenberg magnets: a first example} we will go through the elementary examples of the Heisenberg ferro- and antiferromagnet, which have identical symmetry breaking pattern $SU(2) \to U(1)$ but show completely opposite behavior in their NG mode spectrum and quantum fluctuations. In Section \ref{sec:Quantum fluctuations} I will clarify what I mean by {\em quantum fluctuations}. The next Section \ref{sec:Order parameters and Nambu--Goldstone modes} summarizes the old and recent results on SSB and NG modes. Sections \ref{sec:Order parameters and quantum fluctuations} and \ref{sec:Thin spectrum and decoherence} comprise the main part of this work. In the former I show how one should examine all possible order parameter operators in order to make claims about the presence or absence of quantum fluctuations in an SSB groundstate. The latter explains the intricacies of the thin spectrum of low-lying states and how they influence coherence times of quantum superpositions. It also gives hints as to how the ramifications of the claims made here could be experimentally verified. In Section \ref{sec:Examples} provides a slew of examples of model systems that satisfy properties (i)--(vii) in degrees varying from none to all. A summary and open questions are collected in Section~\ref{sec:Conclusions}.

\section{Heisenberg magnets: a first example}\label{sec:Heisenberg magnets: a first example}
We will illustrate these considerations by the Heisenberg ferromagnet and antiferromagnet, to introduce the general concepts in a concrete system. The 2D square lattice spin-$\tfrac{1}{2}$ antiferromagnet is an exquisite example not only because it has been studied intensively, but mostly because quantum fluctuations are in this case very pronounced: the order parameter, staggered magnetization, in the groundstate with quantum fluctuations being taken into account, is only about 60\% of that of the groundstate of the classical antiferromagnet. In principle a solid like the chair the reader is sitting on is an equally valid example, but there the quantum fluctuations are negligible and it does not resonate with the intuition as much as the antiferromagnet does.

Consider spin-$s$ Heisenberg magnets on bipartite lattices with Hamiltonian
\begin{equation}\label{eq:single exchange Heisenberg Hamiltonian}
 \mathcal{H} = \mathcal{H}_J + \mathcal{H}_\mathrm{ext} = J \sum_{\langle jl \rangle} \mathbf{S}_j \cdot \mathbf{S}_l - \mathbf{h} \cdot \sum_j \mathbf{S}_j.
\end{equation}
The operators $S^a_j, a=x,y,z$ are generators of $SU(2)$ in the spin-$s$ representation at lattice site $j$, $J$ is the exchange parameter, the sum $\langle jl \rangle$ is over nearest-neighbor lattice sites, and $\mathbf{h}$ is an external magnetic field. The commutation relations are
\begin{equation}
 [S^a_j,S^b_l] = \ti \epsilon_{abc} S^c_j \delta_{jl}, \qquad a,b,c = x,y,z.
\end{equation}
The first term in Eq.~\eqref{eq:single exchange Heisenberg Hamiltonian} is invariant under global $SU(2)$-rotations $\te^{\ti \theta^a S^a}$ over angles $\theta^a$ of all spins simultaneously, generated by $S^a = \sum_j S^a_j$. If $J <0$ this term favors alignment of the spins and the groundstate is a ferromagnet. If $J > 0$, the antiferromagnetic case, it favors anti-alignment of the spins and the groundstate  of the corresponding classical Hamiltonian is a N\'eel state with alternating spins.

\subsection{Ferromagnet}\label{subsec:first example ferromagnet}
For the ferromagnetic case $J<0$, if there is no external field $\mathbf{h} = 0$, then alignment in each direction has the same energy. In practice, one of the directions will be chosen {\em spontaneously}. Formally, this can be achieved by imagining a tiny external perturbation $\mathbf{h} >0$ that favors a certain direction, and once the system has reached the ferromagnetic state, let the perturbation vanish $\mathbf{h} \to 0$. Choose the alignment to lie along the $z$-axis, then the order parameter operator is $S^z = \sum_j S^z_j$ itself, which commutes with the Hamiltonian since it is a symmetry generator. Rotations around the $z$-axis, the $U(1)$-group generated by $S^z$, are unbroken, while $S^x$ and $S^y$ are spontaneously broken. According to the adapted Goldstone theorem (see Sec.~\ref{subsec:Nambu--Goldstone modes}), these operators excite the same single NG mode, the magnon, with quadratic dispersion. Concurrently, the magnetization $M = \langle S^z\rangle$ is maximal, such that the raising operator $S^+_j = S^x_j + \ti S^y_j$ would take one out of the Hilbert space, while the lowering operator $S^-_j = S^x_j - \ti S^y_j$ excites the NG mode. In this very special case, the classical groundstate is an eigenstate of the unperturbed Hamiltonian, so this is the whole story. There are no quantum fluctuations and at zero temperature the state will perpetually keep its alignment direction, even for small-size systems. In this sense, the Heisenberg ferromagnet can be called the most classical instance of spontaneous symmetry breaking at zero temperature.

\subsection{Antiferromagnet}\label{subsec:first example antiferromagnet}
In contrast, the antiferromagnet $J>0$ differs strongly from its classical counterpart. In the groundstate of the classical Hamiltonian, the N\'eel state, each site has its spin aligned antiparallel to its nearest neighbors, which is possible without any frustration on bipartite lattices (lattices that can be unambiguously split into two sublattices $A$ and $B$ such that a site at sublattice $A$ only has nearest neighbors on sublattice $B$ and vice versa). In terms of the eigenvalues of $S^z_j$, which can take values in $m_j = -s , -s+1 , \ldots s-1,s$, this state is
\begin{equation}
 \lvert \textrm{N\'eel} \rangle = \otimes_j \lvert m_j = (-1)^j s  \rangle.
\end{equation}
We see that the total state is a (tensor) product of one-particle states. Choose the spin-coordinate system such that the axis of the antialignment is the $z$-axis. Then the order parameter operator quantifying this arrangement is $\mathfrak{S}^z = \sum_j (-1)^j S^z_j$, called the {\em staggered magnetization}, where $(-1)^j$ alternates sign depending on whether $j$ is on the A- or the B-sublattice. Again, $U(1)$-spin rotations around the $z$-axis are unbroken, while $S^x$ and $S^y$ are spontaneously broken. Since $\mathfrak{S}^z$ does not commute with the Hamiltonian, $S^+$ and $S^-$ each excite distinct NG modes, also called magnons, each with linear dispersion relation.

The N\'eel state is obviously not an eigenstate of the Hamiltonian Eq.~\eqref{eq:single exchange Heisenberg Hamiltonian}. The easiest way to see this is to take the raising and lowering operators $S^\pm_j = S^x_j \pm \ti S^y_j$, and rewrite the exchange term of the Hamiltonian as,
\begin{equation}\label{eq:single exchange S+S- Heisenberg Hamiltonian}
  \mathcal{H}_J = J \sum_{\langle jl \rangle} \tfrac{1}{2}( S^+_j S^-_l + S^-_j S^+_l) + S^z_j S^z_l.
\end{equation}
From the first two terms here we see that the Hamiltonian includes processes where the spin on one site is raised while simultaneously that on a neighboring site is lowered. As we argued such processes are forbidden in the ferromagnet, but in the antiferromagnet they occur prominently. Then unitary time evolution $\te^{\ti \mathcal{H}_J t}$ will bring the N\'eel state to a superposition of many other states with a few spins flipped. This can no longer be written as a product of one-particle states. In fact, it has been proven that the groundstate of the antiferromagnetic Hamiltonian Eq.~\eqref{eq:single exchange S+S- Heisenberg Hamiltonian} on any finite lattice must be a total spin singlet, i.e. a totally antisymmetrized superposition of spin states~\cite{Marshall55,LiebMattis62}. But such a state does not break any symmetry (as $\mathbf{S}\lvert\mathrm{singlet}\rangle = 0$, symmetries generated by $\mathbf{S}$ are unbroken). The staggered magnetization of this state vanishes as well. Formally, this singlet is the groundstate for any finite-size system, but in practice many-body systems will be found in a state with broken symmetry and long-range order characterized by staggered magnetization. It was recognized early on by Anderson~\cite{Anderson52} that there is a {\em thin spectrum} or {\em tower of states} of energy vanishing as $1/N$, where $N$ is the number of degrees of freedom, c.q. the number of particles. The actual, symmetry-breaking state is a superposition of the symmetric (singlet) groundstate and states in the the thin spectrum, a wavepacket boasting long-range order, with a very long lifetime. This state corresponds to the spontaneously broken state in the thermodynamic limit. The quantum fluctuations severely modify the classical N\'eel state, tending towards restoration of symmetry. Indeed, already a simple spin-wave theory estimate gives a significant reduction of the staggered magnetization~\cite{Anderson52}, which has been confirmed extensively in numerical simulations~\cite{Manousakis91}. For $s=\tfrac{1}{2}$ and a $D=2$ finite-size square lattice, the staggered magnetization is about 60\% of that of the N\'eel state. The antiferromagnet is a special case in the sense that this strong reduction of symmetry breaking persists even for very large systems.

\section{Quantum fluctuations}\label{sec:Quantum fluctuations}
The main topic of this work is how quantum fluctuations  influence the spontaneously broken groundstate. First we need to define what we mean by quantum fluctuations in this context. For this purpose, we define a `classical state' in a quantum system with SSB. The classical state is then dressed with quantum fluctuations which tend to reduce the amount of symmetry breaking, i.e. reduce the magnitude of the order parameter. Nevertheless in most cases, SSB in the thermodynamic limit will lead to almost precisely this classical state, for instance in superconductors, crystals and most relativistic field theories. However, in some cases the quantum fluctuations cause the groundstate to severely deviate from the classical state, as we have seen in the example of the antiferromagnet in Sec.~\ref{subsec:first example antiferromagnet}. In finite-size systems, due to tunneling between the different degenerate groundstates, there is a unique lowest-energy state which does not break any symmetry at all. This state is however very unstable against perturbations, and there are other states, very close in energy to the absolute groundstate, which do feature broken symmetry. In these cases, when SSB does take place in practice, the `groundstate' realized in nature is a compromise between the classical state and the symmetric groundstate.

For these reasons we identify three states of interest below: the `classical state' $\lvert \Psi_\mathrm{cl} \rangle$ in which symmetry is broken and corresponds to our intuition about SSB; the `quantum groundstate' $\lvert \Psi_\mathrm{symm} \rangle$ which is the unique groundstate in finite-size systems that breaks no symmetry at all; and the `actual groundstate' $\lvert \Psi_0 \rangle$, which is the classical state dressed with quantum fluctuations. It is a very robust state, with broken symmetry. In finite-size systems it is close in energy to $\lvert \Psi_\mathrm{symm} \rangle$ but breaking symmetry related to $\lvert \Psi_\mathrm{cl} \rangle$.

\subsection{Spontaneous symmetry breaking in short}
Let us quickly review the essentials of SSB. Let the Hamiltonian $\mathcal{H}$ be invariant under a group of symmetry transformations $G$. We are only interested in continuous groups $G$ that give rise to NG modes. When the potential is of such a form that the groundstates of this Hamiltonian are only invariant under a subgroup $H$ of $G$, then SSB will take place. There exists a degenerate set of groundstates; given any one of them $\lvert \psi \rangle$, the action of elements $g$ of the coset $G/H$ will lead to a different state $g \rightharpoonup \lvert \psi \rangle \equiv \lvert g \psi \rangle$ which is degenerate in energy since $\mathcal{H}$ commutes with the action of $g$ by definition. Therefore the degenerate groundstates can be labeled by elements of $G/H$. 

The important point is that two inequivalent groundstates $\lvert \psi \rangle$, $\lvert g \psi\rangle$ for $g \neq 1$ are {\em orthogonal} in the thermodynamic limit (see for instance a proof that the matrix elements $\langle \psi \rvert \mathcal{H} \lvert g \psi \rangle$ vanish in that limit in Ref.~\cite{Weinberg96b}). This implies that whenever the system is found in one particular state $\lvert \psi \rangle$, it will stay in that state forever. The Hilbert spaces corresponding to each of the states $\lvert g\psi \rangle$ are disconnected.

One can define an order parameter operator $\mathcal{O}$ that distinguishes the degenerate groundstates, see Eq.~\eqref{eq:SSB definition} below. In all the cases in which we are interested, it is the space integral of a local operator $\mathcal{O} = \int \td^D x\; \mathcal{O}(x)$. Its expectation value takes values in $G/H$, that is,
\begin{equation}\label{eq:groundstate expectation value}
 \langle g \psi \rvert \mathcal{O}(x) \lvert g \psi \rangle = g.
\end{equation}
Here the groundstates are translationally invariant, because the eigenstates of the Hamiltonian are
momentum eigenstates, so the right-hand side is independent of position. This definition holds up to
a multiplicative factor denoting the magnitude of the order parameter. In general states with different
magnitude of the order parameter differ in energy while states connected by a transformation in $G/H$
are degenerate.

Using these notions, we can give the Bogoliubov definition of SSB. Consider the symmetric Hamiltonian $\mathcal{H}$ supplemented with a symmetry breaking field, as parametrized by a chemical potential $\mu$: $\mathcal{H}_\mu = \mathcal{H} - \mu \mathcal{O}$. Any $\mu > 0$ will explicitly break the symmetry by favoring one of the states in $G/H$ over the others, let us call this groundstate $\lvert \psi (N,\mu) \rangle$, where $N$ is the number of particles in a many-body system or the volume in a field theory. The essence of SSB is that taking the thermodynamic limit and the limit of vanishing chemical potential do not commute.
\begin{align}
 \lim_{N \to \infty} \lim_{\mu \to 0} \langle\psi (N,\mu) \rvert\mathcal{O} \lvert\psi (N,\mu) \rangle &= 0, \label{eq:symmetric limits}\\
 \lim_{\mu \to 0} \lim_{N \to \infty} \langle\psi (N,\mu) \rvert \mathcal{O} \lvert\psi (N,\mu) \rangle &\neq 0\label{eq:SSB limits}.
\end{align}
Eq.~\eqref{eq:symmetric limits} is the claim that in the absence of a symmetry-breaking external field there is always a symmetric groundstate. Eq.~\eqref{eq:SSB limits} is called a {\em quasiaverage}. This implies that in the presence of even the slightest perturbation, the broken state is favored over the symmetric state $\lvert \Psi_\mathrm{symm} \rangle$. For finite-size systems, this argument requires only little adjustment. In the presence of an external field, one state is favored. After the field is set to zero, this state persists even though it is not the absolute lowest energy state, because the time to tunnel to that state is extremely long, increasing as $N$ grows. 

\subsection{Three types of groundstates}
Even though one would think that the groundstate is a well-defined physical object, it is precisely the peculiarity of SSB that this notion becomes more complicated. We have already alluded to this in the introduction: the state realized in nature, for instance a periodic crystal, is almost always not an eigenstate of the Hamiltonian. Instead, the system will find itself near its `classical configuration', for instance in a position and not a momentum eigenstate. 

There are several ways to look at the classical state---which is a perfectly quantum state as a vector
in Hilbert space---but there is a very simple and intuitive definition: The classical state is an eigenstate
of the order parameter operator, which extrapolates to the SSB groundstate in the thermodynamic
limit.

In a many-body system, the classical state is the many-body state that can be decomposed a product of single-particle states. These single-particle states are typically generalized coherent states: minimum-uncertainty wavepackets centered around a certain (classical) value, in our case, centered around the order parameter expectation value $\langle \mathcal{O}(x) \rangle$. A good example is the BCS groundstate of a superconductor. For a field theory, the classical state leads to a field configuration which solves the Euler--Lagrange equations of motion.

 There are two reasons for this naming: in many cases we can directly compare classical and quantum field configurations, governed by classical resp. quantum Hamiltonians, and in those cases the groundstates of the classical system map to these states $\lvert \Psi_\mathrm{cl} \rangle$ in the quantum system. All the lattice-spin models considered in this work are examples of that. The second reason is more profound: SSB is a way in which classical behavior emerges from quantum constituents. Everybody would regard a chair or a table as a classical object, while at the heart it is governed by a quantum Hamiltonian. In this regard a superconductor is as classical as a chair, this is precisely the concept of generalized rigidity~\cite{Anderson84}. 

In the limit of vanishing chemical potential, fluctuations come into play. They are the corrections to the classical state due to the dynamical terms in the Hamiltonian or Lagrangian. In the path integral, they are the fluctuations around the classical (stationary) path. The state $\lvert \Psi_0 \rangle$ is the state which is realized in nature, the classical state dressed with/modified by quantum fluctuations. As we have mentioned, in almost all cases, the order parameter operator $\mathcal{O}$ does not commute with the Hamiltonian. The classical state $\lvert \Psi_\mathrm{cl} \rangle$ which is an eigenstate of the order parameter operator, is therefore not an eigenstate of the Hamiltonian. Notwithstanding the different classical states becoming formally orthogonal in the thermodynamic limit, these fluctuations are in principle always present. However, the influence of quantum fluctuations is ``usually utterly negligible'' in the words of Anderson~\cite{Anderson84}. In general, quantum fluctuations are stronger when the number of degrees of freedom is low. Here we are interested not in whether quantum fluctuations have a large or small effect, but in symmetry-broken states that fundamentally have no quantum fluctuations at all. We know this to be the case for the Heisenberg ferromagnet (Sec.~\ref{subsec:first example ferromagnet}), and we want to investigate whether other systems which have an order parameter operator that commutes with the Hamiltonian behave the same way.

In finite-size systems, the different classical states are no longer orthogonal but overlap. Tunneling between the different states $\lvert g \psi \rangle$ is now allowed and will lower the energy. There is a unique groundstate, a certain superposition of all the states $\lvert g \psi \rangle$, which does not break any symmetry at all. This state we call $\lvert \Psi_\mathrm{symm} \rangle$. For instance, in the antiferromagnet, this is the total spin singlet. In finite-size but large systems, SSB can still take place as we know from everyday experience. The reason is that $\lvert \Psi_\mathrm{symm} \rangle$ is inherently unstable against perturbations~\cite{KomaTasaki94,ShimizuMiyadera02}. On the other hand there are states $\lvert \Psi_0 \rangle$ very close in energy to $\lvert \Psi_\mathrm{symm} \rangle$ but with broken symmetry, which extrapolate to the symmetry-broken states in the thermodynamic limit, as has been shown explicitly for $U(1)$-SSB in Ref.~\cite{ KomaTasaki94}.

Summarizing, SSB as is taught in most courses identifies the classical states $\lvert \Psi_\mathrm{cl} \rangle$ as the degenerate, orthogonal groundstates. In principle, there are always quantum fluctuations that lead to a modified state $\lvert \Psi_0 \rangle$. Starting out by assuming there is a long-range ordered state $\lvert \Psi_\mathrm{cl} \rangle$ and consecutively examining the symmetry-restoring quantum fluctuations that lead to $\lvert \Psi_0 \rangle$ is sometimes called the semiclassical method. The effect of quantum fluctuations is habitually rather minimal. The cases where quantum fluctuations are prominent are small systems, but also some large quantum systems like the antiferromagnet of Sec.~\ref{subsec:first example antiferromagnet}. Strictly speaking, the symmetry breaking becomes exact for infinite volume, but for any large, finite size, the antiferromagnet is subject to severe quantum fluctuations.

\subsection{Symmetry breaking in finite systems}
We close this section with some remarks on SSB in finite-size systems. Again, strictly speaking spontaneous symmetry breaking only occurs for infinite systems, or equivalently in the thermodynamic limit. In finite-size systems there exists a unique, symmetric groundstate $\lvert \Psi_\mathrm{symm} \rangle$. It seems that, at least formally, one cannot speak of symmetry breaking of finite systems in the absence of external perturbations. We have the following remarks about this apparent conundrum:
\begin{itemize}
 \item Any system in the universe and especially any system in the laboratory is obviously of finite size. Yet we witness broken symmetries on a daily basis. It would be foolish not to make use of the powerful machinery of order parameters, NG modes, phase transitions etc. while the finite system is really well approximated by the infinite system~\cite{ParkKim10,BirmanNazmitdinovYukalov13}. The broken state in finite systems is directly related to a formally spontaneously-broken state in the thermodynamic limit~\cite{KomaTasaki94}. The validity of this approximation can actually be turned into a quantitative statement, by comparing the time it takes to tunnel from the symmetry-broken state to the groundstate, with the time scale that the observer is interested in. In fact, the thermodynamic limit is for all practical purposes reached very quickly, $N \sim 100 - 10000$.
 \item  The non-commuting limits of Eq.~\eqref{eq:SSB limits} show that any tiny perturbation is sufficient to break the symmetry. No real-world system is completely free of external perturbations. Thus the limiting procedure of maintaining a small external field while taking the thermodynamic limit could be viewed not only as a mathematical trick but also to good approximation as the actual situation in nature as well.
 \item As I will argue later, macroscopic but small systems might well be the most interesting regime for spontaneous symmetry breaking due to the measurable effects of the thin spectrum on the coherence time of quantum superpositions. 
\end{itemize}
For these reasons I will not shy away from using the term ``spontaneous symmetry breaking'' pertaining to finite-size systems. Readers taking a more formal point of view are welcome to replace this with ``explicit symmetry breaking'' by an infinitesimal field and to regard the NG modes having an infinitesimal mass, or take the limiting procedure using external fields literal.

\section{Order parameters and Nambu--Goldstone modes}\label{sec:Order parameters and Nambu--Goldstone modes}
In this section we collect known, formal results about spontaneous symmetry breaking and NG modes. In particular we summarize the recent classification of type-A and type-B NG modes. Soon after the establishment of the Goldstone theorem that connects spontaneously broken symmetries to massless bosonic excitations, it was recognized that it should hold in non-relativistic systems as well, although details need to be modified. In particular, in Lorentz-invariant systems, the dispersion relation of a scalar bosonic mode can only be linear $\omega  \propto k$, since time and space exist on equal footing. In solid state physics it was of course known that the Heisenberg ferromagnet spin wave (magnon) has a quadratic dispersion, and also that there is only one independent spin wave even though there are two broken spin-rotation generators. Lange proved a non-relativistic version of the Goldstone theorem~\cite{Lange65,Lange66} (see also Ref.~\cite{Wagner66}). The precise statement is that there must be an excitation of vanishing energy as momentum goes to zero $k \to 0$ whenever there is a spontaneously broken symmetry. Notably it does not state that each broken generator corresponds to an independent excitation, and it does not specify the dispersion relation. Lange also explicitly distinguishes the NG excitations that appear as $k\to 0$ and are propagating, from those at exactly $k=0$, which ``are just other ground states'' in the degenerate coset space $G/H$. A good early review including remarks on non-relativistic systems can be found in Ref.~\cite{GuralnikHagenKibble68}.

About a decade later, Nielsen and Chadha proved a counting rule, where the number of broken symmetries is equal to or less than the number of NG modes provided that NG modes which have an even-exponent dispersion $\omega  \propto k^{2n}$ are counted twice~\cite{NielsenChadha76}. In the 1990s, Leutwyler established a method for obtaining effective Lagrangians for the low-energy spectrum of non-relativistic systems with broken symmetries~\cite{Leutwyler94}. Here the main ingredient for the distinction between type-A and type-B NG modes is already present, namely the possibility of a term in the effective Lagrangian which is linear in time derivatives, while spatial derivatives must be of at least quadratic order in isotropic systems, such that the dispersion relation $\omega  \propto k^2$ becomes feasible. In the context of kaon condensates in quantum chromodynamics (QCD), it was noted that a NG mode with quadratic dispersion arises and that it is related to the fact that the commutator of a pair of broken generators has a non-vanishing expectation value in the broken groundstate $\langle \Psi_0 | [Q_1,Q_2] | \Psi_0 \rangle \neq 0$~\cite{SchaferEtAl01} (this is formally not correct, see below). Nambu recognized that in this case, two broken symmetry generators in fact excite the same NG mode~\cite{Nambu04}. In various configurations, Brauner, Watanabe and Murayama expanded upon these results and finally proved this statement using effective Lagrangians~\cite{Brauner10,WatanabeBrauner11,WatanabeMurayama12,WatanabeMurayama14r}. The same counting rule was found independently using Mori projector operators by Hidaka~\cite{Hidaka13}. 

\subsection{Spontaneous symmetry breaking}\label{sec:Spontaneous symmetry breaking}
Take a physical model described by a Hamiltonian $\mathcal{H}$ (similar statements can be made for a Lagrangian formalism). Any unitary or antiunitary operator $g$ that commutes with the Hamiltonian $[g,\mathcal{H}]=0$ is said to be a {\em symmetry} of the system. Antiunitary operators can be decomposed into a unitary operator and the time-reversal operator. Since we are interested in continuous symmetries, we shall not discuss the discrete time-reversal operator any further, and consider only unitary transformations. The composition $g_1 g_2$ of two symmetry operators is again a symmetry, all symmetries form a group $G = \{g\}$. If the symmetry is continuous, $G$ is a Lie group. The corresponding Lie algebra $\mathfrak{g}$ is said to be generated by a set $\{Q^a\}$ of {\em symmetry generators}. They are also {\em Noether charges} since they correspond to the space integral of the temporal component of Noether currents $j^a_\mu(x)$,  $Q^a = \int \td^D x\; j_t^a(x)$, which are conserved in time by Noether's theorem. The $j^a(x) \equiv j^a_t(x)$ are called {\em Noether charge densities}, and satisfy Lie algebra equal-time commutation relations,
\begin{equation}\label{eq:Lie algebra relations}
 [j^a(x),j^b(y)] = \ti \sum_c f^{abc} j^c(x) \delta(x-y),
\end{equation}
Here $f^{abc}$ are called {\em the structure constants}, which completely specify the Lie algebra. The charges themselves satisfy $[Q^a,Q^b]  = \ti \sum_c f^{abc} Q^c$.

The group $G$ now consists of elements $g = \exp^{ \ti \theta^a Q^a}$, where $\theta^a$ are real parameters. The Noether charges are observables and therefore Hermitian, so $g$ is unitary. A state $| \psi \rangle$ is said to be invariant under the operation $g$ if $g \rightharpoonup | \psi \rangle = | \psi \rangle$ (since quantum states are defined up to a phase factor, the right-hand side could in principle obtain a phase factor and still be called invariant). Since $g = \exp^{ \ti \theta^a Q^a}$, this requirement is satisfied if $Q^a \rightharpoonup |\psi\rangle =0$, that is, if the Noether charge annihilates the state.

The colloquial way of describing spontaneous symmetry breaking is stating that the groundstate $|\Psi_0\rangle$ is not invariant under all or some symmetries of the Hamiltonian. However, if $Q |\Psi_0\rangle \neq 0$, then this state is actually not well-defined~\cite{Brauner10,GuralnikHagenKibble68}. Namely, the groundstate is assumed to be translationally invariant. One of the reasons for this is that the NG modes are momentum eigenstates, and since $P$ is the generator of translations, good momentum quantum numbers can only be present in states with translational invariance. For large systems, it is sufficient that the translation symmetry be discrete, such that a crystal lattice, which breaks continuous to discrete translations, possesses enough symmetry for NG modes to be well-defined. Because of translational invariance, the norm of the state $Q |\Psi_0\rangle$ would be:
\begin{align}\label{eq:rotated SSB state norm}
 \langle \Psi_0 | Q Q | \Psi_0 \rangle &= \int \td^D x\; \langle \Psi_0 | Q j(x) | \Psi_0 \rangle \nonumber\\
 &=\langle \Psi_0 | Q j(0) | \Psi_0\rangle  \Big( \int \td^D x\;  1 \Big) \to \infty.
\end{align}
Therefore, the formal definition for a symmetry generator $Q^a$ to be spontaneously broken in the groundstate $| \Psi_0 \rangle$ is that there exist an operator $\Phi$ such that,
\begin{equation}\label{eq:SSB definition}
 \langle \Psi_0 | [ Q^a, \Phi ] | \Psi_0 \rangle \neq 0.
\end{equation}
Using the Baker--Campbell--Hausdorff formula,
\begin{equation}\label{eq:Baker--Campbell--Hausdorff formula}
 \te^{X} Y \te^{-X} = Y + [X,Y] + \tfrac{1}{2!} [X,[X,Y]] + \ldots
\end{equation}
it is easy to see that the left-hand side of Eq. \eqref{eq:SSB definition} must vanish if $\te^{\ti \theta^a Q^a}$ leaves the groundstate invariant, so this agrees with our intuition of a broken symmetry. Expectation values of commutators never suffer from the infinities of Eq. \eqref{eq:rotated SSB state norm}. The operator $\Phi$ is called the {\em interpolating operator} (or {\em interpolating field}). The operator $\mathcal{O} \equiv [Q^a,\Phi]$ is called the {\em order parameter operator} and its groundstate expectation value $\langle \mathcal{O} \rangle = \langle \Psi_0 | \mathcal{O} | \Psi_0 \rangle$ is called the {\em order parameter}. Some authors call $\mathcal{O}$ itself the order parameter, but we will always make the distinction. It is a good name, since the order parameter is zero in the disordered, symmetric state while it becomes non-zero in the ordered, symmetry-broken state.

The symmetry generators that are not spontaneously broken form a subgroup $H \subset G$. The broken generators in general do not form a group; instead they form a coset $G/H$. As the broken generators transform the order parameter to a different, degenerate value, the coset space is sometimes called {\em order parameter space}: it enumerates all inequivalent groundstates given an arbitrary reference groundstate. Relative to this reference state,  the value of the order parameter acting on the degenerate broken-symmetry states can be associated uniquely with an element of $G/H$, and can therefore be said to take values in order parameter space. For instance, in Heisenberg magnets, the full symmetry group of spin rotations is $SU(2)$, the group of unbroken symmetries of rotations around the (staggered) magnetization axis is $U(1)$, and the order parameter takes values in $SU(2)/U(1) \simeq S^2$, the two-sphere. Note that our intuitive picture of unit arrows pointing in some direction in three dimensions is precisely this two-sphere.

\subsection{Nambu--Goldstone modes}\label{subsec:Nambu--Goldstone modes}
From Eq. \eqref{eq:SSB definition} one can derive the Goldstone theorem by inserting a complete set of momentum and energy eigenstates, and taking appropriate limits. I will sketch the proof, details can be found for instance in Refs. \cite{GuralnikHagenKibble68,Brauner10}. We start from the definition of a spontaneously broken generator $Q^a$, Eq.~\eqref{eq:SSB definition}. The expectation value $\langle [ Q^a, \Phi ] \rangle \equiv \langle \Psi_0 | [ Q^a, \Phi ] | \Psi_0 \rangle$ is time-independent, because
\begin{align}
 \partial_t \langle [ Q^a (t), \Phi ] \rangle 
 &= \partial_t \int \td^D x\; \langle [ j^a_0 (t,x), \Phi ] \rangle \nonumber\\
 &= -\int \td^D x\; \langle [ \nabla \cdot \mathbf{j}^a (t,x), \Phi ]\rangle \nonumber\\
 &= - \oint \td \mathbf{S} \cdot \langle [ \mathbf{j}^a (t,x), \Phi ]\rangle  =0
\end{align}
Here we use that the Noether current is conserved $\partial_t j^a_0 + \nabla \cdot \mathbf{j}^a =0$ and that the surface term must vanish at infinity. In some cases, the surface term does not vanish and this can lead to interesting physics~\cite{Brauner10,WatanabeMurayama14r}, but we leave this aside here. For the current operator we have $j^a(x,t) = \te^{\ti \mathcal{H} t - \ti P\cdot x} j^a(0) \te^{-\ti \mathcal{H} t + \ti P\cdot x}$ with Hamiltonian $\mathcal{H}$ and momentum operator $P$. We assume that the groundstate is translation invariant and has energy $E_0 = 0$, such that $\te^{-\ti \mathcal{H} t + \ti P\cdot x}\lvert \Psi_0 \rangle =\lvert \Psi_0 \rangle  $.
Now we insert a complete set of states $|n_k \rangle$ with momentum $k$ and energy $E_n(k)$. Then

\begin{align}
 \langle [ Q^a (t), \Phi ] \rangle 
 &= \sum_n \int_k  \int_x \langle \Psi_0 \rvert j^a_0 (t,x) \lvert n_k \rangle \langle n_k \rvert  \Phi \lvert \Psi_0 \rangle \nonumber \\
 &= \sum_n \int_k  \int_x \langle \Psi_0 \rvert j^a_0 (0) \te^{\ti E_n t-\ti kx} \lvert n_k \rangle \langle n_k \rvert  \Phi \lvert \Psi_0 \rangle \nonumber\\
 &\phantom{mmmmmmm} -\langle \Psi_0 \rvert \Phi \lvert n_k \rangle \langle n_k \rvert \te^{-\ti E_n t+\ti kx} j^a_0 (0) \lvert \Psi_0 \rangle  \nonumber\\
 &=  \sum_n \int_k \delta(k) \Big[\te^{\ti E_n t} \langle \Psi_0 \rvert j^a_0 (0) \lvert n_k \rangle \langle n_k \rvert  \Phi \lvert \Psi_0 \rangle  \nonumber\\
 &\phantom{mmmmmmm}  -\te^{-\ti E_n t} \langle \Psi_0 \rvert \Phi \lvert n_k \rangle \langle n_k \rvert j^a_0 (0) \lvert \Psi_0 \rangle  \Big].
\end{align}

Now by Eq.~\eqref{eq:SSB definition} the left-hand side is non-zero. That implies at least one of the terms of the right-hand side must be non-zero. We have also seen that the left-hand side is time-independent, which means that the non-zero terms must have $E_n(k) \to 0$. Since $\int \td^D x\; \te^{\pm i k x} = \delta(k)$, which is peaked around $k \to 0$, we now conclude that there is at least one state $\lvert \pi \rangle \equiv \lvert n_k \rangle$ which is excited by both the Noether charge density $j^a$ and the interpolating field $\Phi$, which has zero energy towards zero momentum, i.e. it is a gapless mode. This is the NG mode. If the theory is Lorentz invariant this mode must have linear dispersion $E_n(k)  \propto k$, but this is clearly not a requirement in the proof of the Goldstone theorem itself.

From this we can already infer that if the order parameter operator is a symmetry generator itself $\mathcal{O} = Q^c$, then via the Lie algebra relations Eq. \eqref{eq:Lie algebra relations} the broken symmetry generators will always come in pairs (they are each other's interpolating field $\Phi$), and via the Goldstone theorem these pairs will excite the same NG mode. It is interesting to note that a relativistic theory can never have an order parameter operator that is a Noether charge density, because a Noether charge density is the temporal component of a four-current, and as such would change value under Lorentz transformations.  There
is an exception: finite Noether charge densities can arise for a Lorentz-invariant Lagrangian when the Lorentz symmetry is broken spontaneously.

Now we state the counting rule as presented by Watanabe and Murayama~\cite{WatanabeMurayama12,WatanabeMurayama14r}. It rests on two observations: the non-vanishing commutator expectation value of broken symmetry generators (the order parameter), and the fact that the term in the effective Lagrangian of NG modes linear in time derivatives is proportional to this order parameter. First define the matrix
\begin{equation}\label{eq:Watanabe-Brauner matrix}
 \rho_{ab} = -\ti \langle \Psi_0 | [ Q^a, j^b(x) ] | \Psi_0 \rangle.
\end{equation}
Because the groundstate is translationally invariant, the matrix is independent of $x$. As a result of the Lie algebra relations Eq. \eqref{eq:Lie algebra relations}, it is a real matrix, and because of the commutator it is antisymmetric. The authors realized that a real, antisymmetric matrix can always be cast in a block diagonal form by a suitable orthogonal transformation, where the blocks are either antisymmetric 2$\times$2-matrices or zero-matrices:
\begin{align}
 \rho_{ab} &\to 
 \begin{pmatrix} 
M_1 & & &  \\
& \ddots & & \\
& & M_K & \\
& & & 0 
 \end{pmatrix},
 \qquad
 M_i = \begin{pmatrix} 0 & \lambda_i \\ -\lambda_i & 0 \end{pmatrix}.
\end{align}
Here the $\lambda_i$ are real and non-zero. The rank of the matrix is: $\mathrm{rank}\; \rho_{ab} = 2K$. The $N$ broken symmetry generators split up into $N-2K$ generators with a separate interpolating field and vanishing commutator expectation value with all other generators, and $K$ pairs with mutual non-vanishing commutator expectation value. Let $\pi^a(x)$ represent the NG field corresponding to $\langle \pi | j^a(x) | 0 \rangle$ in the derivation of the Goldstone theorem. The Fourier transform is $\pi^a(k) = \langle \pi | j^a(k) | 0 \rangle = \langle \pi | \int \td^D x\;\te^{\ti x\cdot k} j^a(x) | 0 \rangle $. Watanabe and Murayama showed that the term in the effective Lagrangian with a single time derivative takes the form (after an orthogonal transformation). 
\begin{equation}\label{eq:single time derivative effective Lagrangian}
 \mathcal{L}^{(1)}_\mathrm{eff} = \frac{1}{4} \rho_{ab} \big( (\partial_t \pi^a) \pi^b - \pi^a (\partial_t \pi^b) \big).
\end{equation}
Thus precisely those generators $Q^a, Q^b$ which have a finite commutator expectation value $\rho_{ab} \neq 0$ will lead to this term, showing that their dynamics are coupled, and the two NG fields correspond to one NG mode. As noted, in general it is an expectation value of a symmetry generator (Noether charge density) $\lambda_i = \langle j^i \rangle$. When this expectation value would vanish, so would this term in the effective Lagrangian. Next, because they feature a single time derivative (besides possible quadratic time derivatives), their dispersion relation is altered. In almost all cases, including all cases of internal symmetry, the lowest order in spatial derivatives is quadratic, leading to a $\omega  \propto k^2$ dispersion relation; some more exotic dispersions have been found when external (spacetime) symmetry is broken~\cite{WatanabeMurayama14r}. For isotropic systems, which are assumed for the spontaneously broken groundstate as is translational symmetry, a term linear in spatial derivatives must vanish. These $K$ modes are called type-B NG modes, while the $N - 2K$ modes without coupled dynamics are called type-A NG modes. In almost all cases, type-A modes have linear, and type-B modes quadratic dispersion. For our purposes here, this shows that an order parameter operator that is a symmetry generator causes a reduced number of quadratically dispersing NG modes. Therefore property (i) implies (ii) (See Sec.~\ref{sec:Introduction}).

A very interesting consequence of this distinction is the effect of fluctuations on the stability of the ordered state in low dimensions. The Mermin--Wagner--Hohenberg--Coleman theorem states that spontaneous fluctuations prevent the formation of an ordered state of infinite size (thermodynamic limit) at dimensions $D \le 2$ for any finite temperature and at dimension $D \le 1$ for zero temperature. The heuristic argument is as follows: the low-energy perturbations of the order parameter are the NG modes. For a stable groundstate, the fluctuations for small distances $x \to 0$ should be small. The correlation function in $D$ spatial dimensions for the order parameter over a distance $x$ involves the propagator of scalar NG modes via
\begin{equation}
 \lim_{x\to 0} \int \td^D k\; \td \omega\; \te^{ikx} \frac{1}{\omega^2 - k^2} \sim \int \td k\; k^{D-1} \frac{1}{k} \sim \int \td k\; k^{D-2}.
\end{equation}
This correlation function has an infrared ($k\to 0$) divergence if $D \le 1$. This is the result for zero temperature; at finite temperature there is an additional factor of $1/k$ from the boson distribution function of the NG modes, which effectively causes the divergence to occur already in one spatial dimension higher. This is in congruence with the notion that a quantum field theory can generally be mapped onto a thermal classical field theory in one dimension higher. However, all these statements assume that the propagator of the NG mode is like $\frac{1}{\omega^2 - k^2}$. If there are only type-B NG modes, we have in fact a different propagator and we need to reevaluate the derivation. Watanabe and Murayama have shown that indeed, if there are only type-B NG modes, there is the possibility of long-range order at zero temperature even in 1+1 dimensions~\cite{WatanabeMurayama14r}. This is corroborated by the established knowledge that long-range order is indeed possible in the ferromagnetic (one-dimensional) spin chain. See Sec.~\ref{sec:Conclusions} for a remark about this statement.

\section{Order parameters and quantum fluctuations}\label{sec:Order parameters and quantum fluctuations}
Now we come to the main part of this work. We investigate how the facts that the order parameter operator is a symmetry generator and the absence of quantum fluctuations, as witnessed in the Heisenberg ferromagnet, are related. In the previous section we have seen that when two spontaneously broken symmetry generators have a non-vanishing groundstate expectation value of their commutator, we get a type-B NG mode. In almost all cases this implies that the order parameter operator is a linear combination of symmetry generators via the Lie algebra relations Eq. \eqref{eq:Lie algebra relations}. The only other option is a so-called central extension of the Lie algebra~\cite{Brauner10}, but here we are not interested in that possibility. Therefore we assume that the order parameter operator is a superposition of symmetry generators, and therefore it commutes with the Hamiltonian.

Many authors, most notably Anderson~\cite{Anderson84}, divide the whole universe of ordered states into those which have an order parameter operator that commutes with the Hamiltonian and those that do not, to immediately afterwards dryly state that the former collection contains only one known example: the Heisenberg ferromagnet. Consequently they tend to overgeneralize the significance of the commuting order parameter operator, and claim that it is a sufficient condition for the absence of quantum fluctuations. The reasoning is as follows: quantum fluctuations are deviations from the classically preferred direction in configuration space of the order parameter effected by unitary time evolution from the classical groundstate $| \Psi_\mathrm{cl}\rangle$. But if the order parameter operator commutes with the Hamiltonian, it is conserved in time, and hence it will not deviate at all. The flaw in this argument is that only the {\em total} order parameter operator $Q = \int \td^D x\; j(x)$ is conserved in time, while the density $j(x)$ is generally not (this is also true for the ferromagnet). Thus in principle, local fluctuations that leave the total order parameter invariant are allowed. Another way of saying this is that there are many eigenstates of the total order parameter operator that have the same eigenvalue, and the Hamiltonian will in general bring one to a mixture of these eigenstates.

\subsection{Finite Noether charge densities}\label{subsec:Finite Noether charge densities}
First we make the following observation concerning the expectation value of symmetry generators. Initially we are concerned about cases where two broken symmetry generators $Q^a$, $Q^b$ have a non-vanishing expectation value of their commutator, such that the order parameter operator is some combination of symmetry generators $ Q^c$, but the symmetry transformations generated by that order parameter operator are themselves unbroken. For simplicity, consider $SU(2)$ where $S^x,S^y$ are broken, and $S^z$ is the order parameter operator, but spin-rotations around the $z$-axis, generated by $S^z$, are themselves unbroken. The other case, where $S^z$ is also broken, as in a canted magnet, could be pictured by starting from the unbroken case but with an additional small breaking of $S^z$ as well. By small we mean the expectation value of the order parameter. This additional broken symmetry will just result in an additional type-A NG mode on top of the type-B mode excited by $S^x$ and $S^y$, see Sec.~\ref{subsec:example Canted magnet}. Larger values of the order parameter can be obtained by adiabatic continuation. The next two sections~\ref{subsec:Highest-weight states}--\ref{subsec:Quantum fluctuations} concern the case of finite Noether charge densities that are themselves unbroken.

In this case the symmetry generator $S^z$ does not annihilate the groundstate, as would be usually the case for symmetry generators. But the groundstate must still be an eigenstate of $S^z$ with finite eigenvalue $M$ since it is an unbroken symmetry generator. Then $\te^{\ti \theta S^z} |\Psi_0 \rangle = \te^{\ti \theta M} | \Psi_0 \rangle$, and this is equivalent to $\lvert \Psi_0 \rangle$ up to a phase factor, which is indistinguishable in quantum mechanics. This expression suffers from the same infinity as in Eq. \eqref{eq:rotated SSB state norm}, but the statement for the Noether charge density $\te^{\ti \theta j^z(x)} |\Psi_0 \rangle = \te^{\ti \theta m} | \Psi_0 \rangle$ is consistent, where $m = M/V$ and $V$ the volume of the system. Note that in a ferromagnet $M$ corresponds to the magnetization which is a perfectly good physical quantity, and which obviously diverges in the thermodynamic limit since it is extensive.

This special case is property (vi) of the enumeration in Sec.~\ref{sec:Introduction}. Since $j^z(x)$ is a Noether charge density which has a non-vanishing groundstate expectation value, it is referred to as a {\em finite Noether charge density}. Here we see that finite Noether charge densities that are unbroken symmetry generators do not annihilate the groundstate but have instead a finite eigenvalue. It is possible to `remove' this finite expectation value by redefining $S^z_\mathrm{new} = S^z_\mathrm{old} - \langle S^z_\mathrm{old} \rangle$ but this modifies the Lie algebra relations Eq.~\eqref{eq:Lie algebra relations} which we do not desire. And since the magnetization is a physically significant quantity, I see no benefit in doing so.

\subsection{Highest-weight states}\label{subsec:Highest-weight states}
The {\em Cartan subalgebra} of a Lie algebra is the subalgebra generated by a maximal set of mutually commuting Lie algebra generators. It has been shown that it is possible to choose the basis of the Cartan subalgebra in such a way that only Noether charge densities that lie in this Cartan subalgebra can obtain a finite groundstate expectation value~\cite{WatanabeBrauner11}. The eigenstates of generators $Q^c$ in the Cartan subalgebra are called {\em weight states} $| \mu \rangle$ with {\em weight} $\mu_c$, collected in a {\em weight vector} $\mu = \{ \mu_c \}$ (see any textbook, e.g. Ref.~\cite{Georgi99}). Thus by definition,
\begin{equation}\label{eq:weight state definition}
 Q^c | \mu \rangle = \mu_c | \mu \rangle.
\end{equation}
Again, one can revert to densities $j^c(x)$ or consider a finite volume here if one is worried about infinities. Then the eigenvalues $\mu_c$ are always finite. It is possible to establish an ordering in the weight vectors $\mu$~\cite{Georgi99}, and the highest one is called the {\em highest-weight vector}, and the state $|\mu\rangle$ is then the {\em highest-weight state}. For finite-dimensional irreducible representations of the Lie algebra, in which we are exclusively interested here, the highest-weight state is unique. This does assume that we have chosen a certain basis for the Lie algebra, since the Cartan subalgebra is unique only up to isomorphism. For instance in a Heisenberg magnet, choosing the Cartan subalgebra to consist of $S^z$, fixes the highest-weight state to be the maximal magnetization in the $z$-direction.

The space of operators not in the Cartan subalgebra is spanned by the {\em root generators} $E_\alpha = \int \td^D x\; e_\alpha(x)$, where $e_\alpha(x)$ is the root generator density, and $\alpha$ is called the {\em root vector}. The root generators change the weight of weight states as $E_\alpha | \mu \rangle \propto | \mu + \alpha\rangle$. An alternative definition of the highest-weight state is that it is annihilated by all positive root generators (the notion of positivity here follows the established ordering, of course). Furthermore $E^\dagger_\alpha = E_{-\alpha}$. Thus the root generators are not Hermitian but the linear combinations $E^+_{\alpha} = \tfrac{1}{\sqrt{2}}(E_\alpha + E_{-\alpha})$ and $E^-_{\alpha} = -\ti \tfrac{1}{\sqrt{2}}(E_\alpha - E_{-\alpha})$ are. These combinations are elements of the Lie algebra of symmetry transformations, are Hermitian and hence observables, and are therefore symmetry generators $Q^a$. The commutation relation is $[E_\alpha,E_{-\alpha}] =  \sum_c \alpha^c Q^c$ and for their densities $[e_\alpha(x),e_{-\alpha}(y)] =  \sum_c \alpha^c j^c(x) \delta(x-y)$ at equal times. For the Hermitian combinations this leads to
\begin{equation}\label{eq:root generator commutation relation}
 [e^+_\alpha(x),e^-_{\alpha}(y)] = \ti \sum_c \alpha^c j^c(x) \delta(x-y).
\end{equation}
For instance in the ferromagnet, $S^z$ spans the Cartan subalgebra, the raising and lowering operators $S^\pm = S^x \pm \ti S^y$ are root generators, $S^+ = (S^-)^\dagger$, and $S^x, S^y$ are symmetry generators. The weight states for spin-$s$ are $\lvert s,m \rangle$, and the root generators act as $S^\pm \lvert s,m \rangle = \sqrt{ s(s+1) - m (m\pm1)} \lvert s, m \pm 1\rangle$.

So far this is textbook Lie algebras. Let us now connect it to finite Noether charge densities. When some $Q^c$ obtains a finite Noether charge density but is itself unbroken, the groundstate is a finite-weight state with $\mu^c = M$ as mentioned above. Each pair of root generators $E_\alpha, E_{-\alpha}$ for which $ \sum_c \alpha^c \langle Q^c \rangle \neq 0$  corresponds to a pair of broken symmetry generators $E^+_\alpha,E^-_\alpha$. They are broken because they do not leave the groundstate invariant, i.e. they change the weight from $\mu$ to $\mu +\alpha$, with at least some $\alpha_c$ non-zero since $ \sum_c \alpha^c \langle Q^c \rangle \neq 0$. The root generators are precisely the combinations that are relevant to NG modes, since they connect to states with a good quantum number for the order parameter $Q^c$. That is, we can derive the Goldstone theorem from Eq. \eqref{eq:root generator commutation relation}. We know from Sec.~\ref{subsec:Nambu--Goldstone modes} that there is only one massless type-B NG mode. I expect but cannot prove in full generality that the root generator that brings the weight closer to zero excites the massless NG mode. For instance, if $\mu_c >0$ and $\alpha_c >0$, then $E_{-\alpha}$ should excite the NG mode. This holds true for all known examples including the ferromagnet and the linear sigma model with chemical potential, see Sec.~\ref{subsec:example Linear sigma model}. 

We are particularly interested in what happens to the other root generator $E_{\alpha}$. First let us assume that we are not in the highest-weight state. Then $E_\alpha |\mu \rangle = |\mu + \alpha\rangle$ is accessible, and there is a mode associated with $| \mu + \alpha , k \rangle = e_\alpha(k) | \mu \rangle$, just as the massless NG mode is associated with $| \mu - \alpha,k \rangle = e_{-\alpha}(k) | \mu \rangle$. Here $e_\alpha (k) = \int \td^D x\; \te^{\ti x \cdot k} e_\alpha(x)$. Since we know that there is only one massless NG mode for the pair $E_\alpha, E_{-\alpha}$, we expect that $| \mu + \alpha , k \rangle$ corresponds to a massive mode, which is a `partner' to the massless NG mode associated with $| \mu - \alpha, k\rangle$. This is completely consistent with the classification scheme of Hayata and Hidaka~\cite{HayataHidaka15}, which identifies ``gapped partner modes'' to a large class of type-B NG modes. We come back to this point in Sec.~\ref{subsec:Multitude of order parameters}. In the explicit example of the linear sigma model with finite chemical potential, see Sec.~\ref{subsec:example Linear sigma model}, these two modes have dispersion relations of the form~\cite{Brauner10},
\begin{equation}\label{eq:typical type-B dispersion}
 \omega_{\pm} = \sqrt{k^2 + \Delta^2} \pm \Delta 
\end{equation}
where $2\Delta$ is the energy gap. The mode $\omega_-$ has indeed a vanishing energy and dispersion $\omega  \propto k^2$ for $k \to 0$ while $\omega_+$ has an energy gap of $2\Delta$. I conjecture that the two modes excited by $E_\alpha, E_{-\alpha}$ will always have such a dispersion relation to quadratic order in $k$. The NG mode excited from a state with zero weight has $\Delta =0$, and states approaching this state, namely states with weight $\mu_c \to 0$, should have similar physical properties. It is then natural to expect that the gap $\Delta$ grows with higher weight, i.e. for a larger order parameter density $m$.


\begin{figure}
\begin{center}
 \includegraphics[width=7cm]{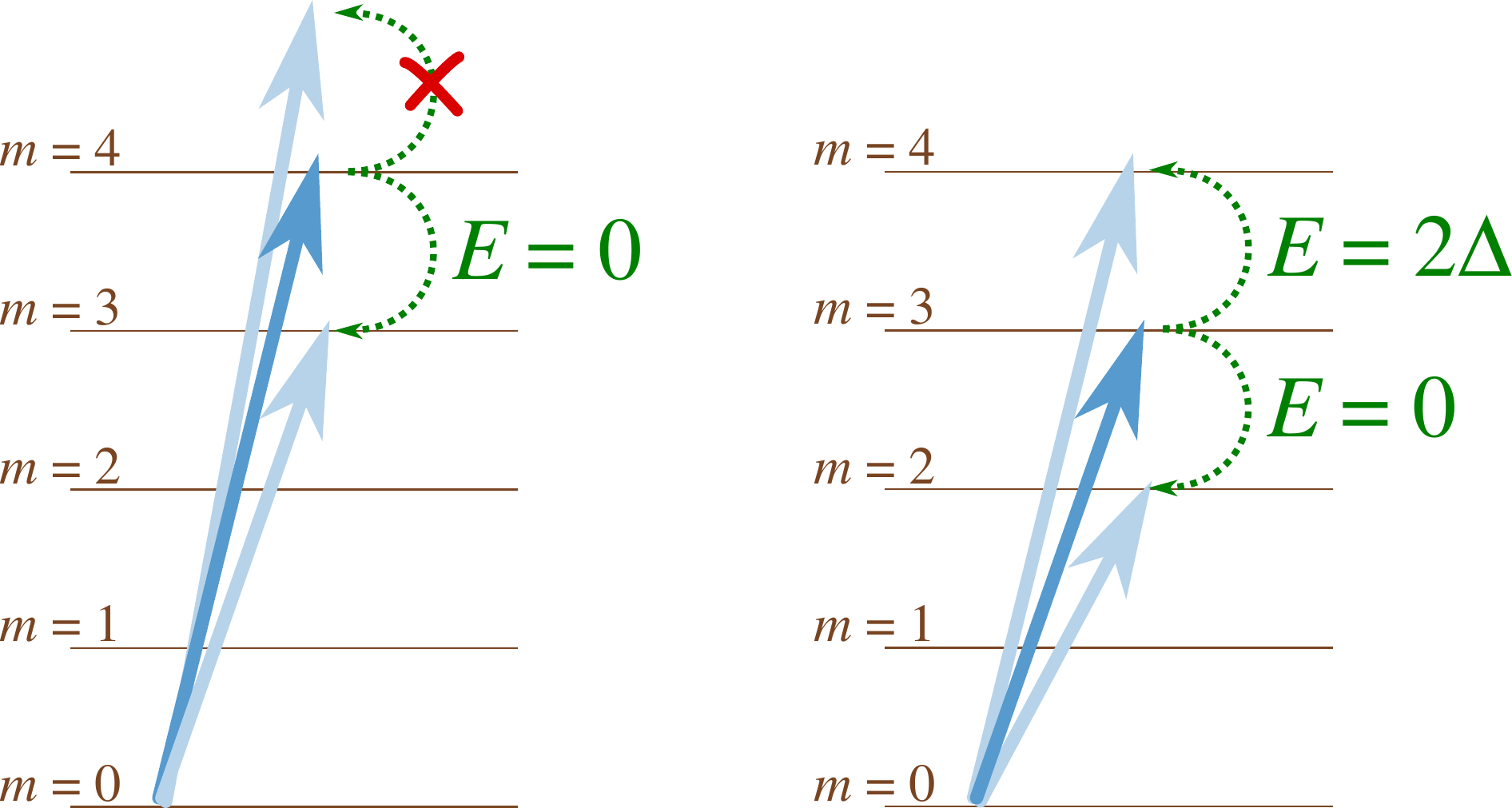}
 \caption{The difference between a fully polarized, ferromagnetic state (left) and an intermediately polarized state (right, see Sec.~\ref{subsec:example Intermediately polarized magnet}) in an $s=4$ Heisenberg magnet. The root generators are the spin raising and lowering operators $S^+$ and $S^-$. In each case, the lowering operator excites a gapless ($E=0$ at $k \to 0$) type-B NG mode. In the intermediately polarized case, the raising operator is able to excite a higher state, and there is an energy gap $2\Delta$ by Eq.~\eqref{eq:typical type-B dispersion}. In the fully polarized case, this operator corresponds to a forbidden operation, and the gapped partner mode is absent. }\label{fig:levels}
\end{center}
 \end{figure}

The other possibility is that we are in the highest-weight state $|\mu \rangle = |\mu_\mathrm{highest}\rangle$. Because of the translational invariance of the groundstate it is the highest-weight state at every point in space. In this case, acting with the positive root $e_\alpha(x)$ at any point $x$ will take one out of the Hilbert space and such operations are therefore forbidden. In this case the second, gapped mode is absent, like in the Heisenberg ferromagnet. The distinction between these two cases is depicted in Fig.~\ref{fig:levels}

\subsection{Quantum fluctuations}\label{subsec:Quantum fluctuations}
We will now show that only groundstates which are highest-weight states cannot have any quantum fluctuations. Recall that here, quantum fluctuations are understood to be the deviations from the classical groundstate $| \Psi_\mathrm{cl} \rangle$ as effected by the quantum Hamiltonian.

The classical groundstate $| \Psi_\mathrm{cl} \rangle$ is an eigenstate of the order parameter operator, with the same eigenvalue at each point in space due to translational invariance. The order parameter takes values in the configuration space which is the coset space $G/H$. The presence of quantum fluctuations means that the groundstate $\lvert \Psi_0 \rangle$ is instead a mixture of states in $G/H$. The only operators that can act on this space are the original symmetry transformations in $G$, generated by the Lie algebra generators $Q^a$. Therefore, the Hamiltonian governing the low-energy excitations consists of compositions of operators $j^a(x)$ corresponding to the densities of the generators $Q^a$. In the previous subsection we have shown how each generator corresponds either to unbroken symmetries, or to broken symmetries which must be Hermitian combinations of root generators.

We again assume that the order parameter operator itself is an unbroken symmetry generator, and that there are no other type-A NG modes  (conversely, if the symmetry generator is broken it excites type-A NG modes and those always lead to quantum fluctuations). In that case, since it commutes with the Hamiltonian, its expectation value is an eigenvalue that is conserved in time. Then the Hamiltonian consists of compositions of generator densities that leave this eigenvalue unchanged. The unbroken symmetries themselves commute with the Hamiltonian and their action cannot change conserved quantities. On the other hand the broken symmetry generator densities are split up into root generators $e_\alpha(x), e_{-\alpha}(x)$. These root generators alter the expectation value of the order parameter as $e_\alpha(x) | \mu , y\rangle \propto \delta(x-y)|\mu +\alpha,y \rangle$, where $| \mu , y\rangle$ is the symmetry-broken groundstate at position $y$. Only compositions of root generators that leave the total order parameter eigenvalue invariant can be allowed. Such terms are for instance of the form $e_\alpha(x) e_{-\alpha}(y)$. They lower the eigenvalue at some point $y$ while raising it by the same amount on point $x$. The terms in the Hamiltonian are compositions of the following three types:
\begin{itemize}
 \item Generators of unbroken symmetries $j^c(x)$;
 \item Generators of broken symmetries at the same location $e_\alpha(x) e_{-\alpha} (x)$;
 \item Generators of broken symmetries at a different location $e_\alpha(x) e_{-\alpha} (y)$ with $x \neq y$.
\end{itemize}
We need to focus only on these three types, compositions of these terms do not alter the argument presented here.

Since the groundstate $| \mu , y\rangle$ is an eigenstate of the first two types, those cannot lead to quantum fluctuations. Acting on the groundstate, these terms give only constant contributions. The dynamics comes from  terms of the form $e_\alpha(x) e_{-\alpha} (y)$. Here we are primarily interested in short-range couplings, since theories with long-range interactions and broken symmetries are more complicated (for instance, gauge bosons mediating long-range interactions are subject to the Higgs mechanism). In a lattice model, short-range couplings are of the form $e_\alpha(x) e_{-\alpha} (x + \delta)$ where $\delta$ is of the order of the lattice constant. In a continuum field theory, these terms are represented by polynomials of spatial gradients $\sim \nabla e_\alpha(x)$. In particular, the NG modes arise from these terms, and since the Goldstone theorem guarantees that NG modes exist whenever the groundstate spontaneously breaks symmetry, terms of the form $e_\alpha(x) e_{-\alpha} (y)$ or compositions thereof are always present.

Summarizing, the allowed terms in the Hamiltonian have the groundstate as an eigenstate, or are of the form $e_\alpha(x) e_{-\alpha}(y)$ with $y\neq x$. It is only the latter type of terms that can bring the classical groundstate to a superposition of states (a mixture of states with the same total eigenvalue of the order parameter). They are always present due to the Goldstone theorem. We see that these terms necessarily involve a factor of the positive root $e_\alpha(x)$. Such terms modify the classical groundstate, {\em unless} they annihilate the groundstate since they would take one outside of the Hilbert space. This is only the case if the groundstate is the highest-weight state. Also in this case, the classical groundstate is an eigenstate of the Hamiltonian, since all terms have either finite eigenvalue or annihilate the state, in other words: have eigenvalue 0. Thus properties (iii) and (iv) are equivalent.

One may now ask: if the dynamic terms annihilate the highest-weight state, how can NG modes exist? The important point is that the NG modes are propagating modes with a finite momentum (even if very small). Furthermore, they are excitations that are caused by perturbations, for instance thermal fluctuations or an external force (linear response). The $k=0$ fluctuations, which are in fact the different classical states in $G/H$, are however always forbidden by the reasoning in this section.

Concluding, we have shown that quantum fluctuations are only absent in groundstates that are highest-weight states, and that they are always present in other cases, even if the order parameter operator commutes with the Hamiltonian. Then total order parameter is conserved, but it may fluctuate locally. Thus (iii) and (iv) always imply (i) but not vice versa. The reverse direction is the topic of the next section.

\subsection{Multitude of order parameters}\label{subsec:Multitude of order parameters}
We now elucidate the structure of order parameter operators necessary to completely specify the SSB consequences. As we have seen above, there exist states with order parameter operators commuting with the Hamiltonian that nevertheless allow for quantum fluctuations. One would like to have a criterion for when and when not such a phenomenon occurs. This is nicely provided by a recent development concerning the full spectrum of excitations related to broken symmetry generators by Hayata and Hidaka~\cite{HayataHidaka15}. These authors consider not just one but all possible order parameter operators. More specifically, they classify all possible interpolating fields collected as
\begin{equation}\label{eq:total interpolating field}
 \mathbf{\Phi}(x) = \begin{pmatrix} \phi^i(x) \\ j^b(x)  \end{pmatrix},
\end{equation}
such that for each component of $\mathbf{\Phi}(x)$ there is at least one symmetry generator $Q^a$ for which $\langle [Q^a , \phi^i(x)]\rangle \neq 0$, or  $\langle [Q^a , j^b(x)]\rangle \neq 0$. Here the $j^b(x)$ are Noether charge densities as before, but the operators $\phi^i$ are different local operators acting on order parameter space $G/H$. The set $\{ \phi^i(x),j^b(x)\}$ should consist of linearly independent operators, and they should excite linearly independent modes from the symmetry-broken groundstate, that is, the determinant of the matrix $\langle [Q^a , \Phi^b(x)]\rangle$ should be non-zero. The Noether charge densities $j^b$ follow the story outlined above, but we also need to consider the other interpolating fields, which will lead to order parameter operators $[Q^a , \phi^i(x)]$ that are not symmetry generators themselves. Because of the transformation properties under the symmetry group, the expectation value $\langle [Q^a,\phi^i] \rangle = \sum_j c_{aij} \langle \phi^j \rangle$  is always some linear combination of the expectation values of the operators $\phi^i$ itself, i.e. not of Noether charge densities $j^b$. An example of the latter is the staggered magnetization of the antiferromagnet (Sec.~\ref{subsec:example Antiferromagnet}) and the ferrimagnet (Sec.~\ref{subsec:example Ferrimagnet}).

Hayata and Hidaka classify the types of NG modes using Eq. \eqref{eq:total interpolating field}. If a symmetry generator $Q^a$ is only broken by interpolating fields $\phi^i$ but not by any Noether charge density $j^b$, it will excite a type-A NG mode. Conversely, if $Q^a$ is broken by a Noether charge density $j^b$, then it will excite a type-B NG mode. So far there is nothing new. However, it may happen that a generator $Q^a$ is broken by both a Noether charge density $j^b$ {\em and} another operator $\phi^i$. In practice, this always involves a pair of symmetry generators $E^+_\alpha, E^-_\alpha$ and a pair of interpolating fields $\phi^i$. The authors have shown that in that case, there is a gapped ``partner'' mode in the spectrum, and this mode is excited by the broken generator $Q^a$, provided the determinant of $\langle [Q^a , \Phi^K(x)]\rangle$ where $K = \{i,b\}$ is non-zero, see \ref{appendix:linear independence of excitations} for a working example. This corresponds precisely to the structure we have found in Sec.~\ref{subsec:Highest-weight states}. We know that one particular combination of the pair of broken Noether charge densities $j^a$, $j^b$, typically the lowering operator $e_{-\alpha}$, excites the massless NG mode. Then the Hermitian conjugate $e_\alpha$ excites the gapped partner mode.

The reason for the presence of quantum fluctuations is now clear: even if the order parameter operator $[Q^a,j^b]$ is conserved in time, the order parameter operator $[Q^a,\phi^i]$ is not. The choice of order parameter always possesses some arbitrariness, since the only requirement is that it be finite in the ordered state while it vanish in the disordered state. We cannot {\em a priori} decide which order parameter is `best', and it befits us to consider all possible operators. Note that a similar course of action is necessary to determine whether a broken symmetry generator $Q^a$ excites a type-A NG mode: one needs to make sure that all the combinations $\langle [Q^a,j^b] \rangle$ in Eq.~\eqref{eq:Watanabe-Brauner matrix} vanish.

The quantum fluctuations could then be said to modify the classical state $\lvert \Psi_\mathrm{cl} \rangle$ as an eigenstate of the order parameter operator $[ Q^a , \phi^i ]$, while leaving it in an eigenstate of the order parameter operator $[ Q^a , j^b ]$. Again, it is clear that the presence of type-A NG modes is associated with order parameters operators $\phi^i$ which do not commute with the Hamiltonian and hence always lead to quantum fluctuations. We can identify the following two cases in the absence of any type-A NG modes:

\begin{enumerate}[1)]
 \item {\em We are in a highest-weight state for all root generators; any fluctuations would take one out of the Hilbert space and are forbidden; there is no additional order parameter operator that does not commute with the Hamiltonian.}
\item {\em There is a finite Noether charge density, but the state is not maximally polarized; the root generators will excite a NG mode and a gapped partner mode; there is an additional order parameter operator not commuting with the Hamiltonian.}
\end{enumerate}

This settles the issue mentioned above: even if the naive order parameter operator---that nevertheless determines the massless NG modes---commutes with the Hamiltonian, there may be other order parameter operators that do not. In that case, quantum fluctuations are present. Note that if one were to look only at the order parameter operators $\phi^i$, one would correctly conclude the spontaneous breaking of symmetry generators $Q^a$, yet one would not recognize that they actually excite type-B and not type-A NG modes. Thus one needs the complete information contained in $ \mathbf{\Phi}(x)$ of Eq. \eqref{eq:total interpolating field} to characterize the low-energy spectrum and the presence of quantum fluctuations.

As an example, consider the ferrimagnet, which is an antiferromagnet of spins of unequal magnitude on each sublattice (see Sec.~\ref{subsec:example Ferrimagnet}). This state has staggered magnetization, but the total magnetization does not cancel out and is finite. In this case, both the magnetization and the staggered magnetization are order parameters and both signal the onset of magnetic ordering in some preferred direction. Since the commutator expectation value of the broken spin rotation generators does not vanish, there is a type-B NG mode which is a spin wave (magnon) of spin precession consistent with the magnetization direction. But since the staggered magnetization is an order parameter as well, there must additionally be a gapped partner mode, the spin wave precession opposite to the magnet, consistent with for instance Refs.~\cite{BrehmerMikeskaYamamoto97,ChubukovEtAl91}.

Another example is the non-maximally polarized ferromagnet, which is explained in more detail in Sec.~\ref{subsec:example Intermediately polarized magnet}. Take a high-spin system $s > 1$ with biquadratic Heisenberg-type Hamiltonian. It could be imagined that for a certain choice of parameters, the groundstate is not a canted state, but a uniform state with lower but non-zero magnetic quantum number $m < s$. This state has a non-zero magnetization like the ordinary, maximally polarized ferromagnet, but quantum fluctuations are possible since the raising operators do not annihilate this state. From a symmetry point of view, the magnetization $M = S^z$ is a good order parameter operator, but the nematic tensor  $N^{zz}$ is as well, where $N^{ab} = \frac{1}{2}(S^a S^b + S^b S^a) - c \delta_{ab}$ and $c$ is some number to make the combination traceless. The nematic tensor does not commute with the Hamiltonian. It can be shown that the excitations generated by  $N^{xz}$ and $N^{yz}$ are linearly independent from those by $S^x,S^y$, and therefore can lead to gapped modes. In the maximally polarized ferromagnet, $N^{zz}$ is also a good order parameter operator, but the excitations are not linearly independent, precluding the existence of a gapped partner mode~\cite{HayataHidaka15,Hidaka14p}. See \ref{appendix:linear independence of excitations} for an explicit derivation.

\subsection{Time-reversal symmetry}\label{subsec:Time-reversal symmetry}
It is claimed by several authors~\cite{Anderson84,Burgess00,Brauner10} that the spontaneous breaking of time-reversal symmetry is responsible for the peculiarities of the ferromagnet, such as the quadratic NG dispersion relation.  It was conceived that the asymmetry between temporal and spatial derivatives in $\omega  \propto k^2$ could only happen when time-reversal symmetry is broken; then a time-reversal odd term in the effective Lagrangian is allowed. This is not the case. Obviously Lorentz invariance is broken in such systems, but there is no fundamental reason time-reversal should be as well. The term with a single time derivative $\sim \pi^a \partial_t \pi^b$ in Eq.~\eqref{eq:single time derivative effective Lagrangian} can be time-reversal even or odd, depending on the transformation properties of $\pi^a$ and $\pi^b$, which follow the properties of the symmetry generators that excite these modes.

Let us take a look at the Lie algebra of symmetry generators and their behavior under time reversal.
Consider the Lie algebra relations $[j^a,j^b]= \ti f^{abc} j^c$, and for simplicity, take the Lie group $SU(2)$ with $f^{abc} = \epsilon^{abc}, \ a,b,c \in x,y,z$. Under time reversal, $\ti \to -\ti$. There are now two possibilities: all three of $j^x,j^y,j^z$ are odd under time-reversal; or two are even and one is odd. The first case is realized in Heisenberg magnets, where the Noether charge densities are spin rotations, which are all odd under time-reversal. In the latter case, a Noether charge density that is even under time-reversal may obtain an expectation value, leading to a type-B NG mode. The groundstate, an eigenstate of the Noether charge density, is then time-reversal invariant as well.  Examples of the second case~\cite{WinklerZulicke10} include the isospin group, which has the same $SU(2)$ structure, but $\tau^x$ and $\tau^z$ are even under time reversal, while $\tau^y$ is odd. If $\tau^z$ were to pick up an expectation value, it would lead to a time-reversal invariant state with a type-B NG mode. The term Eq.~\eqref{eq:single time derivative effective Lagrangian} contains one even field, one odd field and one time derivative, the combination of which is even under time reversal. This was also noticed in Ref.~\cite{WatanabeMurayama14r}. Surely, if a Noether charge density obtains an expectation value $\langle Q^c \rangle = M$, then the $\mathbb{Z}_2$ symmetry $Q^c \to -Q^c$ is spontaneously broken, but this may or may not be the transformation under time-reversal symmetry.

It must be concluded that systems with finite Noether charge densities and type-B NG modes can but not must have spontaneously broken time-reversal invariance, property (vii) in Fig.~\ref{fig:FM equivalencies}.

\section{Thin spectrum and decoherence}\label{sec:Thin spectrum and decoherence}
A very important chapter in the book of spontaneous symmetry breaking concerns the so-called thin spectrum, or (Anderson) tower of states, consisting of a sparse set of states of extremely low energy above the exact groundstate $\lvert \Psi_\mathrm{symm}\rangle$. Its existence was recognized by Anderson~\cite{Anderson52} in 1952 for the Heisenberg antiferromagnet on a square lattice, but its mathematical structure was not realized until later~\cite{HorschVonderlinden88,KaplanLindenHorsch90,KaiserPeschel89,KomaTasaki94}. These states are so close to the groundstate that they are realistically accessible at any finite temperature; the symmetry-broken state realized in nature $|\Psi_0\rangle$ of any finite-size system is then a superposition of the absolute groundstate and states in the thin spectrum. This wavepacket is very stable, and in almost all cases very close to the classical state $|\Psi_\mathrm{cl}\rangle$. It should be regarded at as the groundstate for all practical purposes, for instance in the derivation of the Goldstone theorem (or at least to establish the spectrum of NG modes).

\subsection{Structure of the thin spectrum}
Let us make these statements more precise. True spontaneous symmetry breaking is mathematically rigorous only for infinitely large systems. Then all the states in configuration space $G/H$ are exactly degenerate. After one of these has been spontaneously chosen (perhaps by an external perturbation), it will remain in that state forever, since the tunneling matrix element between the different degenerate states vanishes in the infinite limit~\cite{Weinberg96b}.

For finite systems, a superposition $|\Psi_\mathrm{symm}\rangle$ 
 of all the states in $G/H$ has lower energy than the classical states $| \Psi_\mathrm{cl}\rangle$. This is easy to see starting from a spin-$\tfrac{1}{2}$ antiferromagnet on two sites only: the singlet configuration $\lvert\ua \da\rangle - \lvert\da \ua \rangle$ has a lower energy than any of the pure tensor product (N\'eel) states $\lvert\ua \da\rangle$ or $\lvert\da \ua\rangle$. This extreme case persists for larger systems, and $\lvert\Psi_\mathrm{symm}\rangle$ is the absolute lowest energy state, although the gap to the lowest excitation becomes very small quickly. Koma and Tasaki have shown that the state $|\Psi_\mathrm{symm}\rangle$, which extrapolates to what they call ``naive infinite volume limit'' of the finite-volume groundstate like the spin singlet, is inherently unstable against perturbations~\cite{KomaTasaki94}. 
 
What happens instead is that the would-be degenerate states differ from the groundstate by a tiny energy gap. The spacing of these states goes as $1/N$ where $N$ is the number of degrees of freedom, typically the number of particles. For any macroscopic system the energy of these states is clearly very low, and they become degenerate with the groundstate for $N \to \infty$. If one regards the macroscopic system as a very heavy rotor in configuration space $G/H$ with  moment of inertia $I  \propto N$, then these states are the excitations of this rotor with energy levels like $\hbar^2/I$~\cite{Anderson84,Weinberg96b}. This is the Anderson tower of states. Another name, thin spectrum, was coined to indicate that the density of these states is small, and they do not contribute to the thermodynamic properties of the macroscopic system like specific heat~\cite{KaplanLindenHorsch90}. It can for instance be shown that the contribution to the free energy of a spin system typically scales as $\sim \ln N / N$, which vanishes for large $N$~\cite{KaplanLindenHorsch90,WezelBrink07}.

The notion of the thin spectrum is so profound, that its presence is even used as evidence of long-range order~\cite{BernuLhuillierPierre92}.

\subsection{Thin spectrum as quantum fluctuations}
An alternative view of the thin spectrum was put forward by Van Wezel and coworkers~\cite{WezelBrinkZaanen05,WezelBrinkZaanen06,WezelBrink07}. Since the classical groundstate $|\Psi_\mathrm{cl}\rangle$ is generally not the exact lowest energy state, there are quantum fluctuations tending  towards lower energy. These fluctuations are virtual excitations, which are the deviations of the order parameter from its classically preferred value. As an example, take the spin waves in an antiferromagnet as mentioned in Sec.~\ref{subsec:first example antiferromagnet}. The finite-momentum $k>0$ fluctuations can be readily taken into account, but the ones at exactly zero momentum $k=0$ should be treated separately. In a standard derivation of fluctuations involving Bogoliubov transformations, the $k=0$ components would lead to a singularity~\cite{WezelBrink07,WezelBrinkZaanen06}. These $k=0$ excitations correspond to the fluctuations of the macroscopic, spontaneously broken state {\em as a whole}. They are fluctuations of the center of mass, of the heavy rotor mentioned above. In very small systems, these $k=0$ fluctuations actually occur on a noticeable timescale, and must be properly accounted for~\cite{PapenbrockWeidenmuller14}.

Summarizing, the states in the thin spectrum are the $k=0$ quantum fluctuations of the system as a whole. Then we must immediately conclude that, if a system has no quantum fluctuations as detailed in the previous section, there are also no $k=0$ quantum fluctuations and hence no thin spectrum. This is easy to see in the Heisenberg ferromagnet. The classical, polarized groundstate is an exact eigenstate of the Hamiltonian, and the system as a whole will remain in this state perpetually, even for finite-size systems. There is no lower-energy state to strive towards, and all the classical states are perfectly degenerate. There is no thin spectrum. Therefore property (iii) or equivalently (iv) implies (v) (see Sec.~\ref{sec:Introduction} and Fig.~ \ref{fig:FM equivalencies}).

\subsection{Decoherence time}
The energy of the states in the thin spectrum is lower than that of the lowest NG mode, which has a minimum of momentum inversely proportional to the system size~\cite{WezelBrinkZaanen06}. This causes a fundamental limit on the coherence time of quantum superpositions of small but macroscopic systems. Examples include  superconducting flux qubits, Cooper pair boxes, spins in solids, atomic condensates in optical lattices etc.~\cite{BulutaAshhabNori11}. The argument is as follows~\cite{WezelBrinkZaanen05,WezelBrinkZaanen06}.

We are going to consider a superposition of two slightly different spontaneously broken states, typically one groundstate and one excited state. A good example is two states with $N$ and $N+1$ particles respectively, which is quite literally realized in a Cooper pair box (charge qubit). Call these two states $\lvert m \rangle$ where $m = 0,1$, with energies $E_m$. But in reality, each of them has a thin spectrum of states, labeled by $n$. Therefore we are talking about states $\lvert m,n \rangle$ with energies $E^n_m$. The thin spectrum states have such low energy that they are always thermally populated. Then, we need to consider these thermal mixtures denoted by the density matrix $\rho_m = \frac{1}{Z_m} \sum_n \te^{-\beta E^n_m}  \lvert m ,n\rangle \langle m,n \rvert$. Here $Z_m = \sum_n \te^{-\beta E^n_m}$ is the partition function and $\beta = 1/k_\mathrm{B} T$.

The thermal noise in the states of the thin spectrum is a cause of decoherence which is unavoidable, since these states are so close to the groundstate. It is possible to create an initial superposition of the two states $m = 0,1$ given by the density matrix 
\begin{align}
 \rho_\mathrm{sup} &= \frac{1}{2Z} \sum_n \te^{-\beta E^n_0} \big( \lvert 0,n \rangle \langle 0,n \rvert + \lvert 1,n \rangle \langle 1,n \rvert \nonumber\\
 &\phantom{mmmmmmmmm} +\lvert 0,n \rangle \langle 1,n \rvert +\lvert 1,n \rangle \langle 0,n \rvert \big).
\end{align}
The unitary time evolution via operator $U = \te^{-\ti/\hbar\; \mathcal{H} t}$ of energy eigenstates $\lvert m,n \rangle$ is obviously $U \lvert m,n \rangle = \te^{-\ti/\hbar\; E^n_m t} \lvert m,n \rangle$, and the density matrix evolves as $\rho \to U \rho U^\dagger$. The last two terms in the density matrix pick up phase factors $\te^{\pm\ti/\hbar\; ( E^n_0 - E^n_1) t}$. Since the thin spectrum states are in practice unobservable, we should trace the density matrix over these states. The off-diagonal terms in the density matrix thus obtained are
\begin{equation}
 \rho_\textrm{off-diag} = \frac{1}{2Z} \Big( \sum_n \te^{-\beta E^n_0} \te^{-\ti/\hbar\; ( E^n_0 - E^n_1) t} \Big) \lvert 0 \rangle \langle 1 \rvert
\end{equation}
and its Hermitian conjugate. If these off-diagonal terms vanish, the superposition has been lost: decoherence. Now the phase factors are proportional to $E^n_0 - E^n_1$ which will in general be finite. These phase factors lead to destructive interference, and thereby cause decoherence. The typical timescale for this decoherence is set by the inverse of the energy difference $\Delta E_\mathrm{thin} = E^n_0 - E^n_1$, namely $t_\mathrm{coh}  \propto \hbar / \Delta E_\mathrm{thin}$. Note that the energy difference $E^0_0 - E^0_1$ is constant in time and does not contribute to dephasing (this is the standard constant phase factor for superpositions of states of different energy). The number of states participating in dephasing depends on the temperature, and is of the order of $E_\mathrm{thin} / k_\mathrm{B} T$, where $E_\mathrm{thin}$ is the typical spacing of energy levels in the thin spectrum. Van Wezel and coworkers have shown that in general $\Delta E_\mathrm{thin}  \propto E_\mathrm{thin} /N$~\cite{WezelBrinkZaanen06}. Putting this together, we find a maximum coherence time limited by the dephasing due to thermal population of thin spectrum states of the order of
\begin{equation}\label{eq:thin spectrum coherence time}
 t_\mathrm{coh}  \propto \frac{\hbar}{\Delta E_\mathrm{thin} } \frac{E_\mathrm{thin} }{k_\mathrm{B} T}
 =  \frac{\hbar}{k_\mathrm{B} T} \frac{E_\mathrm{thin} }{\Delta E_\mathrm{thin} } = \frac{\hbar}{k_\mathrm{B} T} N.
\end{equation}

Interestingly, the actual details of the thin spectrum do not matter; because $\Delta E_\mathrm{thin}$ is proportional to $E_\mathrm{thin}$, these factors cancel, and only the size of the system expressed in the number of particles $N$ is of importance.

The maximum coherence time is longer for larger systems, according to Eq.~\eqref{eq:thin spectrum coherence time}. The reason is that the thin spectrum states become more and more degenerate with the groundstate and do not cause dephasing as much. For a life-size system $N \approx 10^{24}$, and at room temperature $\hbar/k_\mathrm{B} T \approx 10^{-14}$s such that $t_\mathrm{coh} \approx 10^{10}$s. But for mesoscopic systems $N \approx 10^{5}$--$10^8$ at laboratory temperature $T \sim 1 K$, this limit $t_\mathrm{coh} \approx 10^{-5}$s starts competing with ordinary, environmental sources of decoherence. 

\subsection{Experiments}
This fundamental limit to coherence time of superpositions of macroscopic systems, which is unavoidable because the thin spectrum is always present and very close to the groundstate (much closer than the lowest `gapless' excitation which has a wavenumber inversely proportional to the system size), is not only of practical importance when creating devices for quantum information purposes, but also provides an experimental test of the claims made in this paper. Namely, in ordered systems where the groundstate is a highest-weight state, like the Heisenberg ferromagnet, there is no thin spectrum and hence this fundamental limit does not apply. 

An experimentalist could craft two very similar systems with a small but macroscopic number of particles, say $N \sim 10000$. For instance the Hamiltonian of one could have as the groundstate $\lvert \Psi_0 \rangle$ a maximally polarized state and the other an intermediately polarized state (see Sec.~\ref{subsec:example Intermediately polarized magnet}). Create macroscopic superpositions, for instance of the groundstate and a magnon state (cf. Ref.~\cite{WezelBrinkZaanen05}). If these systems could be isolated and protected from other sources of decoherence well enough, the system with non-maximal polarization should show decoherence where the maximally polarized one should not. In this way the predictions made in this work could be tested quite directly. 

Note that a real ferromagnet is probably not suitable for such an experiment, since a magnetic material is also a crystal that breaks spacetime symmetry and has phonon NG modes and an associated thin spectrum, next to other sources of decoherence.

\section{Examples}\label{sec:Examples}
\begin{figure}
 \includegraphics[width=1.2cm]{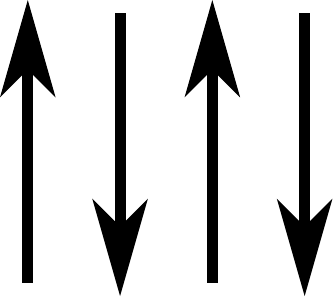}\hfill
 \includegraphics[width=1.2cm]{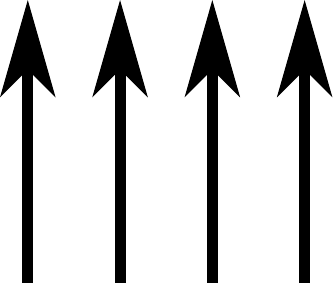}\hfill
 \includegraphics[width=1.2cm]{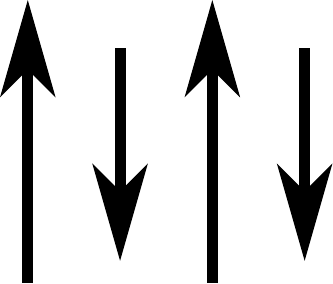}\hfill
 \includegraphics[width=1.2cm]{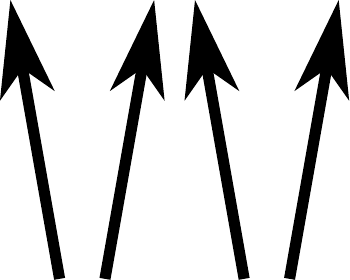}\hfill
 \includegraphics[width=1.2cm]{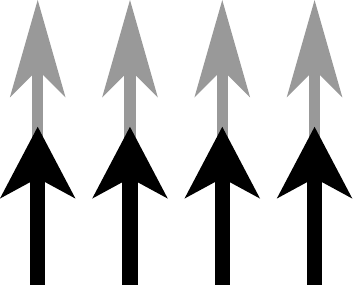}
 \caption{Cartoons of magnetically-ordered phases: antiferromagnet, ferromagnet, ferrimagnet, canted magnet, intermediately-polarized magnet.}\label{fig:magnet cartoons}
\end{figure}

We shall now look at several models that illustrate the SSB, NG mode spectrum and presence of quantum fluctuations in distinct ways. The first few examples are spin-$s$ Heisenberg magnets on bipartite lattices with Hamiltonian
\begin{equation}\label{eq:biquadratic Heisenberg model}
 \mathcal{H} = J \sum_{\langle jl \rangle} \mathbf{S}_j \cdot \mathbf{S}_l + K \sum_{\langle jl \rangle} (\mathbf{S}_j \cdot \mathbf{S}_l )^2
\end{equation}
The operators $S^a_j, a=x,y,z$ are generators of $SU(2)$ in the spin-$s$ representation at lattice site $j$, $J$ is the exchange parameter, $K$ the biquadratic exchange parameter, and the sums are over nearest-neighbor lattice sites. The Hamiltonian is invariant under $SU(2)$-rotations of all spins simultaneously, generated by $S^a = \sum_j S^a_j$. The first term prefers aligned ($J<0$) or anti-aligned ($J>0$) spin on neighboring sites. The biquadratic term, for which we always take $K \ge 0$, prefers to have neighboring spin orthogonally oriented, in the classical case. Thus for small values of $K$ it induces a canting from the aligned or anti-aligned state. In quantum models, this term can also be minimized by choosing a lower magnetic quantum number $m < s$, whereas the ferromagnetic term $J <0$ is minimal for large, aligned values of $m$. See Fig.~\ref{fig:magnet cartoons} for a visual image of the types of magnets considered.

\subsection{Antiferromagnet}\label{subsec:example Antiferromagnet}
Let the exchange parameters be $J>0$, $K=0$. The spins prefer to anti-align, in a spontaneously chosen direction which we can take to be the $z$-axis. The classical groundstate is the N\'eel state, with order parameter operator the staggered magnetization $\mathfrak{S}^z = \sum_j \mathfrak{S}^z_j = \sum_j (-1)^j  S^z_j$. Rotations around the $z$-axis are unbroken, while the other two generators are spontaneously broken: $\langle [S^a,\mathfrak{S}^b_j] \rangle = \ti \epsilon^{abz} \langle \mathfrak{S}^z_j \rangle \neq 0$. The symmetry breaking pattern is $SU(2) \to U(1)$, and the order parameter space is $SU(2)/U(1) \simeq S^2$, the two-sphere. The expectation value of the Noether charge densities $\langle S^z_j \rangle$ is zero. There are two type-A NG modes, magnons with linear dispersion. For low dimension $D$ and low $s$, quantum fluctuations are pronounced, most extreme for $s = \tfrac{1}{2}, D=2$, where the staggered magnetization is reduced to about 60\%~\cite{Manousakis91}. 

\subsection{Ferromagnet}\label{subsec:example Ferromagnet}
Take $J<0$, $K=0$. The groundstate has all spins aligned in a spontaneously chosen direction $z$. The order parameter operator is the generator $S^z$ itself. The symmetry breaking pattern is the same as for the antiferromagnet, $SU(2) \to U(1)$, but now a finite Noether charge density $\langle S^z \rangle = M \neq 0$ is present, such that the two broken densities $S^x_k$ and $S^y_k$ excite the same type-B NG mode with quadratic dispersion. If $M>0$, $S^-_k = S^x_k -\ti S^y_k$ excites the NG mode, while $S^+_k = S^x_k + \ti S^y_k$ takes one out of the Hilbert space. This is case 1) of Sec.~\ref{subsec:Multitude of order parameters} and there is no gapped partner mode. The state is maximally polarized and there are no quantum fluctuations. The classical groundstate is an eigenstate of the Hamiltonian, and there is no thin spectrum, even for small finite-size systems.

\subsection{Ferrimagnet}\label{subsec:example Ferrimagnet}
Ferrimagnets are antiferromagnets that have spins of unequal magnitude on each of the sublattices. We model this with the same Hamiltonian as the antiferromagnet, $J>0$, $K=0$, but the spin $s_A$ on the $A$-sublattice is different from $s_B$ on the $B$-sublattice. For instance, let $A$ have spin-1 and $B$ spin-$\tfrac{1}{2}$. One must be careful about how to formulate the symmetry transformations, but the Hamiltonian has an $SU(2)$-symmetry as before. For $K =0$ we again have the symmetry breaking pattern $SU(2) \to U(1)$ and rotations about the $z$-axis are unbroken. However, because of the imbalance in spin, the expectation value of the generator $S^z$ is non-zero. Thus we have a single type-B NG mode for the broken generators $S^x, S^y$. The staggered magnetization, not commuting with the Hamiltonian, is also an order parameter for this state. We find ourselves in case 2) of Sec.~\ref{subsec:Multitude of order parameters} and there is a gapped partner mode. This is consistent with spin wave theory and numerical calculations~\cite{BrehmerMikeskaYamamoto97,ChubukovEtAl91}.

\subsection{Canted magnet}\label{subsec:example Canted magnet}
Now take a finite biquadratic exchange parameter, $J<0$, $K>|J|/2$. The biquadratic term penalizes a too-high degree of magnetization. This will induce canting of the spins, i.e. classically they will make an angle with the magnetization axis $z$. The canting can be coplanar or not, to be determined by other terms in the Hamiltonian. The classical canting angle $\theta$ is given by $\cos 2\theta = |J|/2s^2K$ for spin-$s$.
The rotation symmetry about the $z$-axis is now spontaneously broken as well, and the symmetry breaking pattern is $SU(2) \to 1$, leading to a type-A NG mode in addition to the type-B NG mode of the ferromagnetic kind. The state is not maximally polarized and there is a gapped partner mode to the type-B mode, as well as the type-A mode. There are quantum fluctuations, for which it is already sufficient to recognize the presence of the type-A NG mode.

It is very instructive to see how the canted magnet emerges from its two limiting cases, the ferromagnet with polarization along the $z$-axis and the antiferromagnet with polarization along the $x$-axis. In the first case, we start out with a ferromagnet with its type-B NG mode excited by $S^x$ and $S^y$ and no gapped partner mode. The expectation value for $S^z$ is maximal, namely $\langle S^z \rangle = M = Ns$. Introducing a canting angle $\theta$ will lower this expectation value to $N (s \cos \theta)$. First of all, excitations that increase the magnetic quantum number, generated by $S^+$, are now possible, and the gapped partner mode emerges. Next, rotations around the $z$-axis are now broken as well. In a coplanar canted magnet, there is now staggered magnetization in for instance the $xz$ plane. This additional breaking of the $U(1)$-symmetry leads to an additional type-A NG mode. Consequently we find one type-A NG mode associated with broken rotations around the magnetization axis, and one type-B NG mode with its gapped partner mode, associated with the other two broken rotations.

Starting out from the antiferromagnet is a bit more intricate. Staggered magnetization along the $x$-axis breaks $S^y$ and $S^z$, and there are two type-A NG modes. Inducing a canting angle leads to a non-zero magnetization in a perpendicular direction, let us say along the $z$-axis. This causes a ferromagnetic-type symmetry breaking of $S^x$ and $S^y$; these two generators now conspire to excite one type-B NG mode and one gapped partner mode. Meanwhile the type-A NG mode excited by $S^z$ persists. We find the same spectrum of one type-A, one type-B and one gapped partner mode. From this point of view there is no difference between a canted ferromagnet and a canted antiferromagnet. These results are consistent with the spectrum of canted magnets derived via effective Lagrangians~\cite{RomanSoto00}.

Note that many canted magnets in nature arise by acting on an antiferromagnet with a magnetic field in another direction than the staggered magnetization axis. This field could be intrinsic or external. In these cases, the additional symmetry breaking is not spontaneous but explicit, i.e. due to a symmetry-breaking term in the Hamiltonian. The new NG mode obtains a gap as a result of this, but the rest of the argument follows similar lines.

\subsection{Intermediately polarized magnet}\label{subsec:example Intermediately polarized magnet}
Again, consider parameters $J<0,K>|J|/2$. For $s > 1$, it is possible that the parameters $J$ and $K$ are fine-tuned such that the average magnetization density $s \cos \theta$ is precisely equal to an allowed magnetic quantum number $s > m > 0$. It could be surmised that the groundstate will have all spins in the state $| s,m\rangle$ (classical spins always have length $s$ and the only thing they can do is canting).
Such intermediately-polarized states are identified in spin-2 and spin-3 spinor BECs, and can be stable in the latter case~\cite{KawaguchiUeda12,TakahashiNitta15}. However, spinor BECs are more complicated since they also carry superfluid sound, see below. 

They follow again the $SU(2) \to U(1)$ pattern, and have finite magnetization $M = Nm$, with $N$ the number of sites. However, the state is clearly not maximally polarized, and $S^-_k$ excites a type-B NG mode and $S^+_k$ a gapped partner mode. See \ref{appendix:Holstein--Primakoff transformation for finite magnetization} for a Holstein--Primakoff derivation of these modes.

The staggered magnetization vanishes, but by the reasoning of Sec.~\ref{subsec:Multitude of order parameters} we know there must be another order parameter operator. In this case the nematic tensor $N^{ab} = \sum_j \tfrac{1}{2} ( S^a_j S^b_j + S^b_j S^a_j) - c \delta_{ab}$ can be seen to obtain an expectation value in the $N^{zz}$-component. The expectation values $\langle [S^x, N^{yz}_j] \rangle$ and $\langle [S^y, N^{xz}_j] \rangle$ do not vanish and indicate the existence of the gapped partner mode. Interestingly, the nematic tensor is also an order parameter operator for the ferromagnet, but then it can be shown that the mode associated with these expectation values is linearly dependent on the NG mode, and cannot be taken as an independent, gapped, mode~\cite{HayataHidaka15,Hidaka14p} (cf.~\cite{TakahashiNitta15}). See \ref{appendix:linear independence of excitations} for more details. 

\subsection{Spinor Bose-Einstein condensates}
Due to the advances in experimental techniques that can cool and trap atoms very efficiently, there has been abundant recent research into Bose--Einstein condensates of atoms with higher spin $s=1,2,3$. These atoms do not only form a bosonic superfluid, but their spin degrees of freedom lead to very interesting physics as well. The symmetry group of the disordered states is $U(1) \times SO(3)$, where the $U(1)$ refers to the superfluid phase variable, and the $SO(3)$ are rotations of the spin degree of freedom. There is a plethora of ordered states which breaks this symmetry to various continuous or discrete subgroups. In particular, a phase with intermediate polarization as described in the previous section is realized in spin-2 under magnetic field (F1 phases) and spin-3 as a groundstate (F phase) spinor-BECs, where we follow the nomenclature of Ref.~\cite{KawaguchiUeda12}.

These systems, also because of their small sizes ($N \sim 10000$), would seem to be ideal testing grounds for the decoherence limit due to the thin spectrum (see Sec.~\ref{sec:Thin spectrum and decoherence}). One complication is that, for certain choices of parameters in the Hamiltonian, the symmetry group may actually be larger than $U(1) \times SO(3)$, leading to additional NG modes. More importantly, one must be very careful to account for the fact that the U(1) superfluid phase is always spontaneously broken, and there is always a type-A NG mode (phonon/zero sound) in the spectrum. This is already sufficient to guarantee the existence of a thin spectrum, which could instigate the same limit even in the spinor-ferromagnetic phase. Whether there is a noticeable difference in decoherence time when there are additional thin spectrum states due to the spin degrees of freedom, is left to further investigation.

\subsection{Linear sigma model}\label{subsec:example Linear sigma model}
As an example from field theory (as opposed to condensed matter physics), consider the linear sigma model for a complex scalar doublet field $\phi = ( \phi_1 , \phi_2)$, with Lagrangian
\begin{equation}\label{eq:linear sigma model}
 \mathcal{L} = \lvert (\partial_t - \ti \mu )\phi \rvert^2 +  \lvert \partial_m \phi \rvert^2 - m^2 |\phi|^2 - \lambda |\phi|^4.
\end{equation}
Here $\mu$ is a chemical potential, not to be confused with the weight vector of Sec.~\ref{subsec:Highest-weight states}. This model captures the physics of the color--flavor-locked phase in kaon condensates in quantum chromodynamics at high density (large $\mu$), and is worked out in detail in Refs.~\cite{SchaferEtAl01,Brauner10}. The Lagrangian without chemical potential is invariant under $SU(2) \times SU(2)$-transformations of the doublet field:
\begin{align}
 \begin{pmatrix} \phi_1 \\ \phi_2 \end{pmatrix} &\to \te^{\ti \vec{\theta}_\mathrm{L} \cdot \vec{\sigma}_\mathrm{L}} \begin{pmatrix} \phi_1 \\ \phi_2 \end{pmatrix},& 
 \begin{pmatrix} \phi_2^* \\ \phi_1 \end{pmatrix} &\to \te^{-\ti \vec{\theta}_\mathrm{R} \cdot \vec{\sigma}_\mathrm{R}} \begin{pmatrix} \phi_2^* \\ \phi_1 \end{pmatrix}.
\end{align}
Here $\vec{\sigma}_{\mathrm{L},\mathrm{R}}$ are vectors containing Pauli matrices. A finite chemical potential  breaks Lorentz invariance, and also breaks the internal symmetry explicitly down to  $SU(2) \times U(1)$. The symmetry generators $\sigma^x_\mathrm{R}$ and $\sigma^y_\mathrm{R}$ are broken explicitly and they excite gapped modes. If $m^2 - \mu^2$ turns negative, the internal symmetry is further broken spontaneously down to the $U(1)$ group generated by $\mathbb{I} + \sigma^z_\mathrm{L}$. The groundstate expectation value of the field can be chosen to be $\langle \phi \rangle \equiv \langle \Psi_0 \rvert \phi \lvert \Psi_0 \rangle = ( 0 , v), v \in \mathbb{R}$; the order parameter operator can be chosen as $\sigma^z_\mathrm{L}$ itself. Therefore a type-B NG boson arises, which is excited by $\sigma^x_\mathrm{L}$ and $\sigma^y_\mathrm{L}$. The symmetry generated by $\mathbb{I} - \sigma^z_\mathrm{L}$ is also spontaneously broken and excites a type-A NG boson. 

The spectrum of the lowest excitations has been worked out in Refs.~\cite{SchaferEtAl01,Brauner10}. The modes excited by $\sigma^x_\mathrm{L}$ and $\sigma^y_\mathrm{L}$ have dispersions $ \omega_\pm  = \sqrt{ k^2 + \mu^2} \pm \mu$. The negative sign corresponds to the NG boson with quadratic dispersion, while the positive sign corresponds to the gapped partner mode with gap $2\mu$. 

We can connect this spectrum to the discussion in Sec.~\ref{subsec:Multitude of order parameters}. From the Lagrangian \eqref{eq:linear sigma model} we can find the Noether charge densities explicitly:
\begin{equation}
 j^A_t = - \ti \pi_a T^A_{ab} \phi_b + \ti \phi^\dagger_a T^A_{ab} \pi^\dagger_b.
\end{equation}
Here we have grouped the symmetry generator matrices as $T^A = ( \mathbb{I} , \sigma^A )$, and the canonical momenta are
\begin{align}\label{eq:linear sigma canonical momenta}
 \pi_a &= \frac{\partial \mathcal{L} }{\partial (\partial_t \phi_a)} = (\partial_t + \ti \mu )\phi^\dagger_a,&
 \pi^\dagger_a &= \frac{\partial \mathcal{L} }{\partial (\partial_t \phi^\dagger_a)} = (\partial_t - \ti \mu )\phi_a.
\end{align}
The canonical commutation relations are
\begin{equation}\label{eq:linear sigma CCR}
 [ \phi_a (x) , \pi_b (y) ] = [ \phi^\dagger_a (x) , \pi^\dagger_b (y) ] = \ti \delta_{ab} \delta(x-y),
\end{equation}
and all other combinations have vanishing commutation relations. The Noether charges are $Q^A = \int \td^D x\; j^A_t(x)$. For the commutation relation between the Noether charges amongst each other one can calculate,
\begin{equation}
 [Q^A , j^B_t (x)]  = - \ti \pi(x) [T^A,T^B] \phi(x) + \ti \phi^\dagger(x) [T^A,T^B] \pi^\dagger(x).
\end{equation}
For our case $\langle \phi \rangle  = ( 0 , v)$, we look at $A,B = x,y$ and $[T^A,T^B] = \ti \sigma^z$. Substituting the canonical momenta Eq.~\eqref{eq:linear sigma canonical momenta}, we find
\begin{align}
 \langle [Q^x,j^y_t(x)] \rangle &= \ti  \langle \big[ - \ti (\partial_t \phi^\dagger) \sigma^z \phi  + \ti \phi^\dagger \sigma^z \partial_t \phi + 2\mu \phi^\dagger \sigma^z \phi \big] \rangle \nonumber\\
 &=\ti 2\mu \big[ \langle \phi^*_1 \phi_1 \rangle - \langle \phi^*_2 \phi_2 \rangle \big] = \ti 2 \mu v^2.
\end{align}
We used the translational invariance of the groundstate to drop the spatial dependence of the left-hand side. Here the groundstate expectation value of the first two terms in the first line vanishes as usual (cf. $U(1)$-symmetry breaking where the Noether charge densities never obtain an expectation value).

In this case, the fields $\phi_1$ and $\phi^\dagger_1$, which do not generate symmetries and are therefore of the form $\phi^i$ in Eq.~\eqref{eq:total interpolating field}, are also interpolating fields for the breaking of $Q^x$ and $Q^y$. We calculate, using Eq.~\eqref{eq:linear sigma CCR},
\begin{align}
 \langle [Q^x, \phi_1(x)] \rangle &= - \ti [\pi_a,\phi_1] T^x_{ab} \langle \phi_b \rangle = - \langle \phi_2 \rangle = -  v,\nonumber\\
 \langle [Q^x, \phi^\dagger_1(x)] \rangle &= \ti \langle \phi^\dagger_a \rangle T^x_{ab}  [\pi^\dagger_b,\phi^\dagger_1]  = \langle \phi^\dagger_2 \rangle =  v,\nonumber\\
 \langle [Q^y, \phi_1(x)] \rangle &= - \ti [\pi_a,\phi_1] T^y_{ab} \langle \phi_b \rangle = \ti \langle \phi_2 \rangle =  \ti v,\nonumber\\
 \langle [Q^y, \phi^\dagger_1(x)] \rangle &= \ti \langle \phi^\dagger_a \rangle T^y_{ab}  [\pi^\dagger_b,\phi^\dagger_1]  = \ti \langle \phi^\dagger_2 \rangle =  \ti v.
\end{align}
For the set $\Phi^A = (j^x_t,j^y_t,\phi_1,\phi^\dagger_1)$ we can check that the excitations are linearly independent by showing that the determinant of the matrix $\langle [Q^A,\Phi^B] \rangle$ is non-zero, in the same way as in \ref{appendix:linear independence of excitations}. Therefore, the type-B NG mode is accompanied by a gapped partner mode, confirming our knowledge of the exact dispersion relations above.

\section{Conclusions}\label{sec:Conclusions}
I have endeavored to provide a satisfactory explanation of why the Heisenberg ferromagnet is such a remarkable state of spontaneously broken matter, even though it is not unique: in the least there are maximally polarized states for any $SU(N)$-system with Heisenberg-type Hamiltonian. The sufficient and necessary condition is that all possible order parameters operators should commute with the Hamiltonian. In the case that there is at least one linearly independent operator, not a symmetry of the Hamiltonian yet obtaining a groundstate expectation value, we find ourselves in an intermediate case. Here the broken generators excite a type-B NG mode {\em and} a gapped partner mode. Furthermore the classical groundstate is not an eigenstate of the Hamiltonian, quantum fluctuations modifying the classical groundstate are present, and a thin spectrum exists. 

These statements can be verified by probing the thin spectrum, which is possible directly in numerical calculations, and indirectly experimentally by investigating the fundamental limit to coherence time of macroscopic superpositions $t_\mathrm{coh}  \propto  \frac{\hbar}{k_\mathrm{B} T} N$. 

Let us conclude with some open questions. First, as mentioned above, spinor BECs appear to be an appealing playground to find type-B NG modes and maximally or intermediately polarized states. The gapped partner mode in the latter case should be identifiable. However, spinor BECs are a superfluid as well with spontaneously broken $U(1)$-symmetry leading to a type-A NG mode in the form of superfluid zero sound. The order parameter associated with this symmetry breaking is of the ordinary kind, and leads to quantum fluctuations and a thin spectrum. As such, a clear-cut experiment that compares two situations with and without a thin spectrum is impossible here. However, the introduction of additional thin spectrum states by going from a maximally to non-maximally polarized state could lead to a significant {\em difference} in coherence time. A careful consideration of the arguments laid out in Refs.~\cite{WezelBrinkZaanen05,WezelBrinkZaanen06} supported by numerical calculations or simulations could provide an answer.

Another question is the stability of the groundstate due to quantum fluctuations. In Ref.~\cite{WatanabeMurayama14r} it is argued that systems which only have type-B and no type-A NG modes, it is possible to have long-range order at zero temperature even in 1+1 dimensions. They provide two derivations based on effective Lagrangians, but the heart of the argument is that there are no quantum fluctuations that can destroy the ordered state: a Heisenberg ferromagnet is stable even at small sizes. However, here we have seen that if the groundstate is not a highest-weight state, quantum fluctuations are in fact present, even if there are only type-B NG modes. It would be interesting to investigate whether long-range order can in fact persist in such 1+1-dimensional systems, and if not, where the argument using effective Lagrangians fails.

Since we have seen in Sec.~\ref{subsec:Time-reversal symmetry} that time-reversal is actually a separate issue, the opposite question becomes relevant: are there examples of `ordinary' symmetry breaking with only type-A NG modes which have nevertheless broken time-reversal symmetry? As a matter of fact, the N\'eel state in an antiferromagnet breaks this symmetry, so the question becomes really: to what extent is broken time-reversal symmetry a relevant feature at all? The standard argument, put forward for instance in Ref.~\cite{Burgess00} is that, in the N\'eel state, the combined transformation of time reversal and translation by one lattice spacing does leave this state invariant. But a combined transformation of time reversal and 180$^\circ$ rotation of all spins also leaves the groundstate invariant, and this symmetry operation is identical for ferromagnets and antiferromagnets alike. It seems there is demand for a more rigorous definition of ``macroscopic order parameter'' that somehow averages over small length scales. Using the improved definition, perhaps it can then be shown that spontaneously broken time-reversal symmetry can only emerge for systems containing type-B NG modes.

\section*{Acknowledgements}
I thank Haruki Watanabe, Yoshimasa Hidaka, Jasper van Wezel, Louk Rademaker, Tomas Brauner and Naoto Nagaosa for useful discussions. This work was supported by the Foreign Postdoctoral Researcher program at RIKEN.

\appendix
\section{Linear independence of excitations}\label{appendix:linear independence of excitations}
Here we state and illustrate when Noether charge densities and other operators as interpolating fields excite linearly independent modes. We should consider the set of all possible interpolating fields in Eq. \eqref{eq:total interpolating field}, namely both the Noether charge densities $j^b$ and other fields $\phi^i$. Then the set of broken Noether charge densities $\{Q^a\}$ has a non-vanishing commutator expectation value for at least one component of $\langle [Q^a,j^b] \rangle$ or $\langle [Q^a,\phi^i] \rangle$. For each of these non-vanishing combinations one could apply the Goldstone theorem, leading to massless modes. However some of these modes may in fact be linearly dependent. 

The criterion is the following~\cite{HayataHidaka15,Hidaka14p}. Take the set of interpolating fields  $\Phi^K = \{ \phi^i,j^b \}$ where $K$ runs over all $\phi^i$ and $j^b$ that causes a symmetry generator to be spontaneously broken. Compute all the commutator expectation values $\mathcal{M}_{KL} = \int \td^D x' \langle \Psi_0 | [\Phi^K(x'),\Phi^L(x) ] |\Psi_0 \rangle$, where as always $|\Psi_0\rangle$ is the symmetry-breaking groundstate. If the determinant of the matrix $\mathcal{M}_{KL}$ is zero, then some of the modes are linearly dependent.

As an example, take a $s=\tfrac{3}{2}$ Heisenberg magnet. The spin matrices in this representation are
\begin{align}
 S^x &= \frac{1}{2} \begin{pmatrix} & \sqrt{3} & & \\ \sqrt{3} & & 2 & \\ & 2 & & \sqrt{3} \\ & & \sqrt{3} & \end{pmatrix},
 \end{align}
 \begin{align}
 S^y &= \frac{\ti}{2} \begin{pmatrix} & -\sqrt{3}  & & \\ \sqrt{3} & & -2 & \\ & 2 & & -\sqrt{3} \\ & & \sqrt{3} & \end{pmatrix},
 \end{align}
 \begin{align}
 S^z &= \begin{pmatrix} \tfrac{3}{2}&  & & \\ & \tfrac{1}{2} & & \\ & & -\tfrac{1}{2} & \\ & &   & -\tfrac{3}{2} \end{pmatrix}.
\end{align}
The nematic tensor is $N^{ab} = \frac{1}{2}(S^aS^b + S^bS^a) - \frac{5}{4} \delta_{ab}$. It is symmetric in $(ab)$ and specified by six components,

\begin{align}
N^{xx} &= \frac{1}{2} \begin{pmatrix} -1 & &  \sqrt{3}  & \\ & 1 & & \sqrt{3} \\ \sqrt{3} & & 1 &  \\ & \sqrt{3} &  & -1 \end{pmatrix}, & 
N^{xy} &= \frac{\ti}{2} \begin{pmatrix}  & &  -\sqrt{3}  & \\ & & & -\sqrt{3} \\ \sqrt{3} & &  &  \\ & \sqrt{3} &  &  \end{pmatrix},  \nonumber\\ 
  N^{yy} &= \frac{1}{2} \begin{pmatrix} -1 & &  -\sqrt{3}  & \\ & 1 & & -\sqrt{3} \\ -\sqrt{3} & & 1 &  \\ & -\sqrt{3} &  & -1 \end{pmatrix}, &
   N^{xz} &= \frac{1}{2} \begin{pmatrix}  & \sqrt{3} &   & \\ \sqrt{3} & & &  \\  & &  &-\sqrt{3}  \\ & &-\sqrt{3}   &  \end{pmatrix}, \nonumber\\ 
   N^{zz} &= \frac{1}{2} \begin{pmatrix} 1 & & & \\ & -1 & & \\ & & -1 &  \\ & &  & 1 \end{pmatrix},  &
 N^{yz} &= \frac{\ti}{2} \begin{pmatrix}  & - \sqrt{3} &   & \\ \sqrt{3} & & &  \\  & &  &\sqrt{3}  \\ & &-\sqrt{3}   &  \end{pmatrix}.
\end{align}

Suppose that the groundstate is the intermediately polarized state $| \Psi_0 \rangle = \begin{pmatrix} 0 & 1 & 0 & 0 \end{pmatrix}^\mathrm{T}$. Both $S^z$ and $N^{zz}$ are order parameters for this state ($N^{xx}$ and $N^{yy}$ are as well, but they do not lead to additional breaking of symmetry generators, so that the former two suffice). The interpolating fields for the spontaneous breaking of $S^x$ are $S^y$ and $N^{yz}$ while those for the spontaneous breaking of $S^y$ are $S^x$ and $N^{xz}$. To determine whether the interpolating fields excite modes linearly independent from the NG mode excited by the pair $S^x,S^y$, we should evaluate the matrix
\begin{equation}\mathcal{M} =
 \begin{pmatrix} 
  0 & \langle [S^x,S^y] \rangle & \langle [S^x,N^{xz}] \rangle &  \langle [S^x,N^{yz}] \rangle \\
   \langle [S^y,S^x] \rangle & 0  & \langle [S^y,N^{xz}] \rangle &  \langle [S^y,N^{yz}] \rangle \\
   \langle [N^{xz},S^x] \rangle & \langle [N^{xz},S^y] \rangle & 0 & \langle [N^{xz},N^{yz}] \rangle \\
   \langle [N^{yz},S^x] \rangle & \langle [N^{yz},S^y] \rangle & \langle [N^{yz},N^{xz}] \rangle & 0
 \end{pmatrix}
\end{equation}
For our groundstate $| \Psi_0 \rangle = \begin{pmatrix} 0 & 1 & 0 & 0 \end{pmatrix}^\mathrm{T}$ this matrix is
\begin{equation}
 \frac{\ti}{2}
 \begin{pmatrix} 
  0 & 1 & 0 & -3 \\
  -1 & 0 & 3 & 0 \\
  0 & -3 & 0 & -3 \\
  -3 & 0 & 3 & 0 
 \end{pmatrix},
\end{equation}
and its determinant is $9 \neq 0$. So here we have linearly independent modes and this indicates the existence of a gapped partner mode, which is excited by the spin raising operator $S^+$. Conversely, suppose that the groundstate is the maximally polarized state $| \Psi_0 \rangle = \begin{pmatrix} 1 & 0 & 0 & 0 \end{pmatrix}^\mathrm{T}$. In this case, we have the same structure for the broken symmetry generators, interpolating fields and order parameters. But the above matrix evaluates to
\begin{equation}
 \frac{\ti}{2}
 \begin{pmatrix} 
  0 & 3 & 0 & 3 \\
  -3 & 0 & -3 & 0 \\
  0 & 3 & 0 & 3 \\
  -3 & 0 & -3 & 0 
 \end{pmatrix},
\end{equation}
and its determinant is zero. In this case, the modes excited by the interpolating fields $N^{xz}, N^{yz}$ are actually linearly dependent on the NG mode, and there is no separate gapped partner mode. We know that the reason is that the raising operator would take one out of the Hilbert space.

\section{Holstein--Primakoff transformation for finite magnetization}\label{appendix:Holstein--Primakoff transformation for finite magnetization}
We want to examine excitations around the state of intermediate magnetization $s > m > 0$. To this end we perform a very simplistic Holstein--Primakoff-type calculation to get a sense of the important issues. We assume that the biquadratic term $\sim K$ serves to stabilize this groundstate, but that the spin waves are primarily due to the ordinary exchange term $\sim J$. In other words, we neglect the biquadratic term here. There are excitations that lower the spin and those that raise the spin. Therefore we introduce quanta $c^\dagger$ that raise the expectation value of $S^z$ and quanta $d^\dagger$ that lower this expectation value. Both excitations have the state $\lvert m \rangle \equiv \lvert s,m \rangle$ as the groundstate: $c \lvert m \rangle = d \lvert m \rangle = 0$. Since the magnetization $m$ is finite, the species do not have the same weight within the decomposition. 

Introduce boson operators $c_i,c_i^\dagger$ and $d_i,d_i^\dagger$ such that $c_i$ and $d_i$ annihilate the state $| m \rangle$ on each site $i$. The commutation relations are $[c_i,c_j^\dagger] = [d_i,d_j^\dagger] = \delta_{ij}$ and the other ones vanish.  Define:
\begin{align}
S^z &= m + c^\dagger c - d^\dagger d , \\
S^+ &= \tfrac{1}{\sqrt{2}} ( \alpha c^\dagger U + \sqrt{\alpha^2 + 1}\; U d),\\
S^- &= \tfrac{1}{\sqrt{2}} ( \sqrt{\alpha^2 + 1}\; d^\dagger U +  \alpha U c),\\
U &= \sqrt{2m - \frac{2}{\alpha^2 - 1} c^\dagger c -\frac{2}{\alpha^2 + 2} d^\dagger d}.
\end{align}
Here $\alpha$ should be a positive real number unequal to 1. The commutation relations $[S^z , S^\pm] = \pm S^\pm$ are easily checked. The remaining one is
\begin{align}
 [S^+ , S^-] &= \tfrac{1}{2} [\alpha c^\dagger U + \sqrt{\alpha^2 + 1}\; U d,\sqrt{\alpha^2 + 1}\; d^\dagger U +  \alpha U ] \nonumber\\
 &= \tfrac{1}{2} \alpha^2 [c^\dagger U,Uc] + \tfrac{1}{2} (\alpha^2+1) [Ud,d^\dagger U] \nonumber\\
 & \phantom{m} + \tfrac{1}{2} \alpha\sqrt{\alpha^2+1} ([ c^\dagger U,Ud] + [ d^\dagger U,Uc] ).
\end{align}
The first two terms do reduce to $2S^z$. The last two terms are hard to calculate because of the square root in $U$. For now we assume they do vanish, they should not matter to lowest order anyway. For the lowest excitations, the radical $U$ is unimportant, and we approximate $U \approx \sqrt{2m}$.

The Heisenberg Hamiltonian is
\begin{equation}
\mathcal{H} = J \sum_{j \delta} \tfrac{1}{2} S^+_j S^-_{j+\delta} + \tfrac{1}{2} S^-_j S^+_{j+\delta} + S^z_j S^z_{j+\delta}.
\end{equation}
Here $j$ runs over all sites and $\delta$ over nearest-neighbors of $j$. We substitute the expressions above and keep only terms up to quadratic order in $c,d$. 
\begin{align}
 \mathcal{H}&= J \sum_{j \delta}\tfrac{m}{2} ( \alpha c^\dagger_j +\sqrt{\alpha^2 + 1}\; d_j)(\sqrt{\alpha^2 + 1}\; d^\dagger_{j +\delta} + \alpha c_{j+\delta}) \nonumber\\
 &\phantom{mmm}+\tfrac{m}{2} (\sqrt{\alpha^2 + 1}\; d^\dagger_{j} + \alpha c_{j})( \alpha c^\dagger_{j+\delta} +\sqrt{\alpha^2 + 1}\; d_{j + \delta} ) \nonumber\\
 &\phantom{mmm}  + m^2 + m c^\dagger_j c_{j} + m c^\dagger_{j+\delta}c_{j+\delta} - m d^\dagger_j d_{j} - m d^\dagger_{j+\delta}d_{j+\delta} \nonumber\\
 &= E_0 + J m\sum_{j \delta}  \tfrac{1}{2} \alpha^2 (c^\dagger_j c_{j+\delta} + c^\dagger_{j+\delta}c_{j}) 
 + \tfrac{1}{2} (\alpha^2+1) (d^\dagger_j d_{j+\delta} + d^\dagger_{j+\delta}d_{j}) \nonumber\\
 &\phantom{mmm} + \tfrac{1}{2} \alpha \sqrt{\alpha^2 +1} (c^\dagger_j d^\dagger_{j +\delta} + d^\dagger_{j}c^\dagger_{j+\delta} + c_{j}d_{j + \delta} + d_j c_{j+\delta}) \nonumber\\
 &\phantom{mmm} + c^\dagger_j c_{j} +  c^\dagger_{j+\delta}c_{j+\delta} -  d^\dagger_j d_{j} -  d^\dagger_{j+\delta}d_{j+\delta} 
\end{align}

Here $E_0 = J m^2Nz$. We Fourier transform in the usual way to find the factors $\nu_k = \sum_\delta \te^{\ti k \cdot \delta}$, which leads to
\begin{align}
 \mathcal{H}&= E_0 + J m\sum_k (\alpha^2 \nu_k +2) c^\dagger_k c_k + ((\alpha^2+1) \nu_k -2) d^\dagger_k d_k \nonumber\\
 &\phantom{mmmmmmmm} + \alpha \sqrt{\alpha^2 +1} \nu_k (c^\dagger_k d^\dagger_k + c_k d_k).
\end{align}
Now we perform a Bogoliubov transformation
\begin{align}
 c_k &= \sqrt{\alpha^2 +1}\; a_k - \alpha b^\dagger_k,& 
 c^\dagger_k &= \sqrt{\alpha^2 +1}\; a^\dagger_k - \alpha b_k, \nonumber\\
 d_k &= -\alpha a^\dagger_k + \sqrt{\alpha^2 +1}\; b_k, &
 d^\dagger_k &= -\alpha a_k + \sqrt{\alpha^2 +1}\; b^\dagger_k
\end{align}
This transformation makes the cross terms $a^\dagger b^\dagger$ and $ab$ vanish, and the Hamiltonian is diagonalized. The Hamiltonian reduces to
\begin{equation}
\mathcal{H} = E_0 + J m\sum_k 2a^\dagger_k a_k + (-1+\nu_k) b^\dagger_k b_k.
\end{equation}
Since $\nu_k = \cos(k\delta) \approx (1 - \delta^2 k^2)$, where $\delta$ is the lattice constant, we see that the $b^\dagger_k$ excite a gapless mode with quadratic dispersion. The $a^\dagger_k$ excitations are gapped, with the gap size set by the exchange parameter $J$ and the magnetization $m$. This mode is actually dispersive if higher order terms are taken into account, but the gap at zero momentum is $Jm$. Now we can identify the modes $b^\dagger_k$ and $a^\dagger_k$. Looking at the Bogoliubov transformations, $b^\dagger_k$ consists of $d^\dagger_k$ and $c_k$, which make up $S^-$, while $a^\dagger_k$ consists of $d_k$ and $c^\dagger_k$ which make up $S^+$. Therefore, for positive magnetization $m$, the lowering operator $S^-_k$ excites the gapless Goldstone mode, while the raising operator $S^+_k$ excited the gapped partner mode.

\bibliographystyle{model1a-num-names}
\bibliography{FMreferences2}

\begin{thebibliography}{46}
\expandafter\ifx\csname natexlab\endcsname\relax\def\natexlab#1{#1}\fi
\providecommand{\bibinfo}[2]{#2}
\ifx\xfnm\relax \def\xfnm[#1]{\unskip,\space#1}\fi
\bibitem[{Anderson et~al.(1990)Anderson, Langacker, and
  Mann}]{AndersonLangackerMann90}
\bibinfo{author}{P.~W. Anderson}, \bibinfo{author}{P.~Langacker},
  \bibinfo{author}{A.~K. Mann}, \bibinfo{journal}{Phys. Today}
  \bibinfo{volume}{43} (\bibinfo{year}{1990}) \bibinfo{pages}{117}.
\bibitem[{Peierls et~al.(1991)Peierls, Kaplan, and
  Anderson}]{PeierlsKaplanAnderson91}
\bibinfo{author}{R.~Peierls}, \bibinfo{author}{T.~A. Kaplan},
  \bibinfo{author}{P.~W. Anderson}, \bibinfo{journal}{Phys. Today}
  \bibinfo{volume}{44} (\bibinfo{year}{1991}) \bibinfo{pages}{13}.
\bibitem[{Lange(1965)}]{Lange65}
\bibinfo{author}{R.~V. Lange}, \bibinfo{journal}{Phys. Rev. Lett.}
  \bibinfo{volume}{14} (\bibinfo{year}{1965}) \bibinfo{pages}{3--6}.
\bibitem[{Lange(1966)}]{Lange66}
\bibinfo{author}{R.~V. Lange}, \bibinfo{journal}{Phys. Rev.}
  \bibinfo{volume}{146} (\bibinfo{year}{1966}) \bibinfo{pages}{301--303}.
\bibitem[{Wagner(1966)}]{Wagner66}
\bibinfo{author}{H.~Wagner}, \bibinfo{journal}{Zeit. Phys.}
  \bibinfo{volume}{195} (\bibinfo{year}{1966}) \bibinfo{pages}{273--299}.
\bibitem[{Nielsen and Chadha(1976)}]{NielsenChadha76}
\bibinfo{author}{H.~B. Nielsen}, \bibinfo{author}{S.~Chadha},
  \bibinfo{journal}{Nucl. Phys. B} \bibinfo{volume}{105} (\bibinfo{year}{1976})
  \bibinfo{pages}{445}.
\bibitem[{Sch\"afer et~al.(2001)Sch\"afer, Son, Stephanov, Toublan, and
  Verbaarschot}]{SchaferEtAl01}
\bibinfo{author}{T.~Sch\"afer}, \bibinfo{author}{D.~Son},
  \bibinfo{author}{M.~Stephanov}, \bibinfo{author}{D.~Toublan},
  \bibinfo{author}{J.~Verbaarschot}, \bibinfo{journal}{Phys. Lett. B}
  \bibinfo{volume}{522} (\bibinfo{year}{2001}) \bibinfo{pages}{67--75}.
\bibitem[{Nambu(2004)}]{Nambu04}
\bibinfo{author}{Y.~Nambu}, \bibinfo{journal}{J. Stat. Phys.}
  \bibinfo{volume}{115} (\bibinfo{year}{2004}) \bibinfo{pages}{7--17}.
\bibitem[{Brauner(2010)}]{Brauner10}
\bibinfo{author}{T.~Brauner}, \bibinfo{journal}{Symmetry} \bibinfo{volume}{2}
  (\bibinfo{year}{2010}) \bibinfo{pages}{609--657}.
\bibitem[{Watanabe and Brauner(2011)}]{WatanabeBrauner11}
\bibinfo{author}{H.~Watanabe}, \bibinfo{author}{T.~Brauner},
  \bibinfo{journal}{Phys. Rev. D} \bibinfo{volume}{84} (\bibinfo{year}{2011})
  \bibinfo{pages}{125013}.
\bibitem[{Watanabe and Murayama(2012)}]{WatanabeMurayama12}
\bibinfo{author}{H.~Watanabe}, \bibinfo{author}{H.~Murayama},
  \bibinfo{journal}{Phys. Rev. Lett.} \bibinfo{volume}{108}
  (\bibinfo{year}{2012}) \bibinfo{pages}{251602}.
\bibitem[{Hidaka(2013)}]{Hidaka13}
\bibinfo{author}{Y.~Hidaka}, \bibinfo{journal}{Phys. Rev. Lett.}
  \bibinfo{volume}{110} (\bibinfo{year}{2013}) \bibinfo{pages}{091601}.
\bibitem[{Watanabe and Murayama(2014)}]{WatanabeMurayama14r}
\bibinfo{author}{H.~Watanabe}, \bibinfo{author}{H.~Murayama},
  \bibinfo{journal}{Phys. Rev. X} \bibinfo{volume}{4} (\bibinfo{year}{2014})
  \bibinfo{pages}{031057}.
\bibitem[{Kawaguchi and Ueda(2012)}]{KawaguchiUeda12}
\bibinfo{author}{Y.~Kawaguchi}, \bibinfo{author}{M.~Ueda},
  \bibinfo{journal}{Phys. Rep.} \bibinfo{volume}{520} (\bibinfo{year}{2012})
  \bibinfo{pages}{253--381}.
\bibitem[{Watanabe and Murayama(2013)}]{WatanabeMurayama13}
\bibinfo{author}{H.~Watanabe}, \bibinfo{author}{H.~Murayama},
  \bibinfo{journal}{Phys. Rev. Lett.} \bibinfo{volume}{110}
  (\bibinfo{year}{2013}) \bibinfo{pages}{181601}.
\bibitem[{Anderson(1984)}]{Anderson84}
\bibinfo{author}{P.~W. Anderson}, \bibinfo{title}{Basic notions of condensed
  matter physics}, volume~\bibinfo{volume}{55} of
  \textit{\bibinfo{series}{Frontiers in Physics}},
  \bibinfo{publisher}{Benjamin/Cummings Pub. Co.}, \bibinfo{year}{1984}.
\bibitem[{Weinberg(1996)}]{Weinberg96b}
\bibinfo{author}{S.~Weinberg}, \bibinfo{title}{The Quantum Theory of Fields},
  volume~\bibinfo{volume}{2}, \bibinfo{publisher}{Cambridge University Press},
  \bibinfo{year}{1996}.
\bibitem[{van Wezel and van~den Brink(2007)}]{WezelBrink07}
\bibinfo{author}{J.~van Wezel}, \bibinfo{author}{J.~van~den Brink},
  \bibinfo{journal}{Am. J. Phys.} \bibinfo{volume}{75} (\bibinfo{year}{2007})
  \bibinfo{pages}{635--638}.
\bibitem[{Hayata and Hidaka(2015)}]{HayataHidaka15}
\bibinfo{author}{T.~Hayata}, \bibinfo{author}{Y.~Hidaka},
  \bibinfo{journal}{Phys. Rev. D} \bibinfo{volume}{91} (\bibinfo{year}{2015})
  \bibinfo{pages}{056006}.
\bibitem[{Gongyo and Karasawa(2014)}]{GongyoKarasawa14}
\bibinfo{author}{S.~Gongyo}, \bibinfo{author}{S.~Karasawa},
  \bibinfo{journal}{Phys. Rev. D} \bibinfo{volume}{90} (\bibinfo{year}{2014})
  \bibinfo{pages}{085014}.
\bibitem[{Marshall(1955)}]{Marshall55}
\bibinfo{author}{W.~Marshall}, \bibinfo{journal}{Proc. Roy. Soc. Lond.}
  \bibinfo{volume}{A232} (\bibinfo{year}{1955}) \bibinfo{pages}{48--68}.
\bibitem[{{Lieb} and {Mattis}(1962)}]{LiebMattis62}
\bibinfo{author}{E.~{Lieb}}, \bibinfo{author}{D.~{Mattis}},
  \bibinfo{journal}{J. Math. Phys.} \bibinfo{volume}{3} (\bibinfo{year}{1962})
  \bibinfo{pages}{749--751}.
\bibitem[{Anderson(1952)}]{Anderson52}
\bibinfo{author}{P.~W. Anderson}, \bibinfo{journal}{Phys. Rev.}
  \bibinfo{volume}{86} (\bibinfo{year}{1952}) \bibinfo{pages}{694--701}.
\bibitem[{Manousakis(1991)}]{Manousakis91}
\bibinfo{author}{E.~Manousakis}, \bibinfo{journal}{Rev. Mod. Phys.}
  \bibinfo{volume}{63} (\bibinfo{year}{1991}) \bibinfo{pages}{1--62}.
\bibitem[{{Koma} and {Tasaki}(1994)}]{KomaTasaki94}
\bibinfo{author}{T.~{Koma}}, \bibinfo{author}{H.~{Tasaki}},
  \bibinfo{journal}{J. Stat. Phys.} \bibinfo{volume}{76} (\bibinfo{year}{1994})
  \bibinfo{pages}{745--803}.
\bibitem[{Shimizu and Miyadera(2002)}]{ShimizuMiyadera02}
\bibinfo{author}{A.~Shimizu}, \bibinfo{author}{T.~Miyadera},
  \bibinfo{journal}{Phys. Rev. Lett.} \bibinfo{volume}{89}
  (\bibinfo{year}{2002}) \bibinfo{pages}{270403}.
\bibitem[{Park and Kim(2010)}]{ParkKim10}
\bibinfo{author}{J.-H. Park}, \bibinfo{author}{S.-W. Kim},
  \bibinfo{journal}{Phys. Rev. A} \bibinfo{volume}{81} (\bibinfo{year}{2010})
  \bibinfo{pages}{063636}.
\bibitem[{Birman et~al.(2013)Birman, Nazmitdinov, and
  Yukalov}]{BirmanNazmitdinovYukalov13}
\bibinfo{author}{J.~Birman}, \bibinfo{author}{R.~Nazmitdinov},
  \bibinfo{author}{V.~Yukalov}, \bibinfo{journal}{Phys. Rep.}
  \bibinfo{volume}{526} (\bibinfo{year}{2013}) \bibinfo{pages}{1 -- 91}.
\bibitem[{Guralnik et~al.(1968)Guralnik, Hagen, and
  Kibble}]{GuralnikHagenKibble68}
\bibinfo{author}{G.~Guralnik}, \bibinfo{author}{C.~Hagen},
  \bibinfo{author}{T.~Kibble}, in: \bibinfo{booktitle}{Advances in Particle
  Physics}, volume~\bibinfo{volume}{2}, \bibinfo{publisher}{Interscience},
  \bibinfo{address}{New York}, \bibinfo{year}{1968}, pp.
  \bibinfo{pages}{567--708}.
\bibitem[{Leutwyler(1994)}]{Leutwyler94}
\bibinfo{author}{H.~Leutwyler}, \bibinfo{journal}{Phys. Rev. D}
  \bibinfo{volume}{49} (\bibinfo{year}{1994}) \bibinfo{pages}{3033--3043}.
\bibitem[{Georgi(1999)}]{Georgi99}
\bibinfo{author}{H.~M. Georgi}, \bibinfo{title}{{Lie algebras in particle
  physics; 2nd ed.}}, Frontiers in Physics, \bibinfo{publisher}{Perseus},
  \bibinfo{address}{Cambridge}, \bibinfo{year}{1999}.
\bibitem[{Brehmer et~al.(1997)Brehmer, Mikeska, and
  Yamamoto}]{BrehmerMikeskaYamamoto97}
\bibinfo{author}{S.~Brehmer}, \bibinfo{author}{H.-J. Mikeska},
  \bibinfo{author}{S.~Yamamoto}, \bibinfo{journal}{J. Phys.: Condens. Matter}
  \bibinfo{volume}{9} (\bibinfo{year}{1997}) \bibinfo{pages}{3921}.
\bibitem[{Chubukov et~al.(1991)Chubukov, Ivanova, Ivanov, and
  Korutcheva}]{ChubukovEtAl91}
\bibinfo{author}{A.~V. Chubukov}, \bibinfo{author}{K.~I. Ivanova},
  \bibinfo{author}{P.~C. Ivanov}, \bibinfo{author}{E.~R. Korutcheva},
  \bibinfo{journal}{J. Phys.: Condens. Matter} \bibinfo{volume}{3}
  (\bibinfo{year}{1991}) \bibinfo{pages}{2665}.
\bibitem[{Hidaka(2014)}]{Hidaka14p}
\bibinfo{author}{Y.~Hidaka}, \bibinfo{howpublished}{private communication},
  \bibinfo{year}{2014}.
\bibitem[{Burgess(2000)}]{Burgess00}
\bibinfo{author}{C.~Burgess}, \bibinfo{journal}{Phys. Rep.}
  \bibinfo{volume}{330} (\bibinfo{year}{2000}) \bibinfo{pages}{193 -- 261}.
\bibitem[{Winkler and Z\"ulicke(2010)}]{WinklerZulicke10}
\bibinfo{author}{R.~Winkler}, \bibinfo{author}{U.~Z\"ulicke},
  \bibinfo{journal}{Phys. Lett. A} \bibinfo{volume}{374} (\bibinfo{year}{2010})
  \bibinfo{pages}{4003--4006}.
\bibitem[{{Horsch} and {Linden}(1988)}]{HorschVonderlinden88}
\bibinfo{author}{P.~{Horsch}}, \bibinfo{author}{W.~v.~d. {Linden}},
  \bibinfo{journal}{Z. Phys. B} \bibinfo{volume}{72} (\bibinfo{year}{1988})
  \bibinfo{pages}{181--193}.
\bibitem[{Kaplan et~al.(1990)Kaplan, von~der Linden, and
  Horsch}]{KaplanLindenHorsch90}
\bibinfo{author}{T.~A. Kaplan}, \bibinfo{author}{W.~von~der Linden},
  \bibinfo{author}{P.~Horsch}, \bibinfo{journal}{Phys. Rev. B}
  \bibinfo{volume}{42} (\bibinfo{year}{1990}) \bibinfo{pages}{4663--4669}.
\bibitem[{Kaiser and Peschel(1989)}]{KaiserPeschel89}
\bibinfo{author}{C.~Kaiser}, \bibinfo{author}{I.~Peschel}, \bibinfo{journal}{J.
  Phys. A} \bibinfo{volume}{22} (\bibinfo{year}{1989}) \bibinfo{pages}{4257}.
\bibitem[{Bernu et~al.(1992)Bernu, Lhuillier, and
  Pierre}]{BernuLhuillierPierre92}
\bibinfo{author}{B.~Bernu}, \bibinfo{author}{C.~Lhuillier},
  \bibinfo{author}{L.~Pierre}, \bibinfo{journal}{Phys. Rev. Lett.}
  \bibinfo{volume}{69} (\bibinfo{year}{1992}) \bibinfo{pages}{2590--2593}.
\bibitem[{van Wezel et~al.(2005)van Wezel, van~den Brink, and
  Zaanen}]{WezelBrinkZaanen05}
\bibinfo{author}{J.~van Wezel}, \bibinfo{author}{J.~van~den Brink},
  \bibinfo{author}{J.~Zaanen}, \bibinfo{journal}{Phys. Rev. Lett.}
  \bibinfo{volume}{94} (\bibinfo{year}{2005}) \bibinfo{pages}{230401}.
\bibitem[{van Wezel et~al.(2006)van Wezel, Zaanen, and van~den
  Brink}]{WezelBrinkZaanen06}
\bibinfo{author}{J.~van Wezel}, \bibinfo{author}{J.~Zaanen},
  \bibinfo{author}{J.~van~den Brink}, \bibinfo{journal}{Phys. Rev. B}
  \bibinfo{volume}{74} (\bibinfo{year}{2006}) \bibinfo{pages}{094430}.
\bibitem[{Papenbrock and Weidenm\"uller(2014)}]{PapenbrockWeidenmuller14}
\bibinfo{author}{T.~Papenbrock}, \bibinfo{author}{H.~A. Weidenm\"uller},
  \bibinfo{journal}{Phys. Rev. C} \bibinfo{volume}{89} (\bibinfo{year}{2014})
  \bibinfo{pages}{014334}.
\bibitem[{Buluta et~al.(2011)Buluta, Ashhab, and Nori}]{BulutaAshhabNori11}
\bibinfo{author}{I.~Buluta}, \bibinfo{author}{S.~Ashhab},
  \bibinfo{author}{F.~Nori}, \bibinfo{journal}{Rep. Prog. Phys.}
  \bibinfo{volume}{74} (\bibinfo{year}{2011}) \bibinfo{pages}{104401}.
\bibitem[{Rom\'an and Soto(2000)}]{RomanSoto00}
\bibinfo{author}{J.~M. Rom\'an}, \bibinfo{author}{J.~Soto},
  \bibinfo{journal}{Phys. Rev. B} \bibinfo{volume}{62} (\bibinfo{year}{2000})
  \bibinfo{pages}{3300--3315}.
\bibitem[{Takahashi and Nitta(2015)}]{TakahashiNitta15}
\bibinfo{author}{D.~A. Takahashi}, \bibinfo{author}{M.~Nitta},
  \bibinfo{journal}{Ann. Phys. (N.Y.)} \bibinfo{volume}{354}
  (\bibinfo{year}{2015}) \bibinfo{pages}{101--156}.

\end{thebibliography}
\end{document}